\documentclass[10pt, journal, final]{IEEEtran}
\IEEEoverridecommandlockouts
\usepackage{bbm}
\usepackage{amsmath}
\usepackage{amsfonts}
\usepackage[pdftex]{graphicx}
\usepackage{subfigure}

\usepackage{times}
\usepackage{cite}
\usepackage{array}
\usepackage{mathrsfs}
\usepackage{multirow}
\usepackage{amssymb}

\usepackage{stfloats}
\usepackage{footnote}
\usepackage{booktabs}

\usepackage{subeqnarray}
\usepackage{cases}
\usepackage{threeparttable}
\usepackage{xcolor}
\usepackage{hyperref}
\usepackage{algpseudocode}
\usepackage{bm}
\usepackage{adjustbox}

\usepackage[linesnumbered,ruled,vlined]{algorithm2e} 
\usepackage{algpseudocode}

\usepackage{threeparttable}
\usepackage{url}
\usepackage{colortbl}
\usepackage{array}

\usepackage[absolute]{textpos}
\def\BibTeX{{\rm B\kern-.05em{\sc i\kern-.025em b}\kern-.08em
		T\kern-.1667em\lower.7ex\hbox{E}\kern-.125emX}}

\begin{document}

\begin{textblock}{18}(2,0.5)
    \centering
    \noindent X. Zhang, S. Sun, M. Tao, Q. Huang, and X. Tang, “Multi-Satellite Cooperative Networks: Joint Hybrid Beamforming and User Scheduling Design,” accepted to \textit{IEEE Transactions on Wireless Communications}, 2023.
\end{textblock}

\title{Multi-Satellite Cooperative Networks: Joint Hybrid Beamforming and User Scheduling Design}

\author{\IEEEauthorblockN{Xuan Zhang, \IEEEmembership{Student Member, IEEE},
Shu Sun, \IEEEmembership{Member, IEEE},
Meixia Tao, \IEEEmembership{Fellow, IEEE}, \\
Qin Huang, \IEEEmembership{Senior Member, IEEE}, and
Xiaohu Tang, \IEEEmembership{Senior Member, IEEE}}

\thanks{This paper was accepted for presentation in part at IEEE Wireless Communications and Networking Conference (WCNC) 2023 \cite{wcnc2023}.}
\thanks{This work is supported by the NSF of China under grants 62271310 and 61941106, and by the Fundamental Research Funds for the Central Universities of China. (Corresponding authors: Shu Sun, Meixia Tao.)}
\thanks{Xuan Zhang, Shu Sun, and Meixia Tao are with the Department of Electronic Engineering and the Cooperative Medianet Innovation Center (CMIC), Shanghai Jiao Tong University, Shanghai 200240, China (e-mail: \{zhangxuan1998, shusun, mxtao\}@sjtu.edu.cn).}
\thanks{Qin Huang is with the School of Electronic and Information Engineering, Beihang University, Beijing 100191, China (e-mail: qhuang.smash@gmail.com).}
\thanks{Xiaohu Tang is with the School of Information Science and Technology, Southwest Jiaotong University, Chengdu, Sichuan 611756, China (e-mail: xhutang@swjtu.edu.cn).}

}

\maketitle
\vspace{-1cm}

\begin{abstract}
In this paper, we consider a cooperative communication network where multiple low-Earth-orbit (LEO) satellites provide services to multiple ground users (GUs) cooperatively at the same time and on the same frequency. 
The multi-satellite cooperation has great potential in extending communication coverage and increasing spectral efficiency. 
Considering that the on-board radio-frequency circuit resources and computation resources on each satellite are restricted, we aim to propose a low-complexity yet efficient multi-satellite cooperative transmission framework. 
Specifically, we first propose a hybrid beamforming method consisting of analog beamforming for beam alignment and digital beamforming for interference mitigation. 
Then, to establish appropriate connections between the satellites and GUs, we propose a heuristic user scheduling algorithm which determines the connections according to the total spectral efficiency increment of the multi-satellite cooperative network. 
Next, considering the intrinsic connection between beamforming and user scheduling, a joint hybrid beamforming and user scheduling (JHU) scheme is proposed to dramatically improve the performance of the multi-satellite cooperative network. 
In addition to the single-connection scenario, we also consider the multi-connection case using the JHU scheme. 
Extensive simulations conducted over different LEO satellite constellations and across various GU locations demonstrate the superiority of the proposed schemes in both overall and per-user spectral efficiencies.
\end{abstract}

% Keywords

\IEEEpeerreviewmaketitle

\section{Introduction}
In the past decades, with the evolution from the first generation mobile communication to the fifth generation (5G), terrestrial wireless communication has experienced unprecedented rapid development, providing people with dense coverage, considerable transmission rate, and affordable services \cite{What5G}.
However, there is still a large area on the earth that is deficient in coverage by terrestrial mobile networks, such as underdeveloped areas, remote villages, maritime and polar regions.
Moreover, in recent years, the demand for multimedia services such as video services is increasing rapidly. The huge requirement for ubiquitous coverage and efficient transmission brings challenges to the development of mobile communication. These challenges can not be addressed by terrestrial mobile networks alone because of the diversified service types and the uncertain distribution of ground users (GUs). Under these circumstances, satellite communication is a feasible and promising solution to share the communication burden of terrestrial systems.

Satellites have a distinctive ability to cover wide geographical areas through a minimum amount of ground infrastructure, which enables them to be an attractive solution to fulfill the growing diversified applications and services, either as a stand-alone system or as an integrated satellite-terrestrial network for beyond-5G and 6G wireless communications \cite{challenge, NGSO}. 
Since their inception, satellite communications have found a great quantity of applications, including media broadcasting, backhauling, news gathering, and so on.
Currently, the field of satellite communications is drawing increasing attention in the global telecommunications market. Several network operators have started using satellites in backhaul infrastructures for connectivity and for 5G system integration \cite{5G}. 
Following the evolution of Internet-based applications, satellite communications are going through a transformation phase refocusing the system design on data services, namely broadband satellite communications. 
The main motivation is a) the rapid adoption of media streaming instead of linear media broadcasting and b) the urgent need to extend broadband coverage to unserved and underserved areas \cite{SatCom}.
Furthermore, satellites are likely to play an increasingly important role in the 6G era to provide global seamless coverage and support space-terrestrial integrated networks \cite{space}. 
The integrated architectures, applications, and challenges of satellite-terrestrial networks toward 6G were presented in \cite{integrated-6G}.
In \cite{Security}, the authors introduced several satellite communication networks which are categorized by different architectures, including land mobile satellite communication networks, hybrid satellite-terrestrial relay networks, and satellite-terrestrial integrated networks. 
The authors in \cite{leo-comparison} made a technical comparison of three low-Earth-orbit (LEO) satellite constellation systems: OneWeb's, SpaceX's, and Telesat's. 
The advantages and technical challenges of multi-satellite cooperative transmission systems in 5G were discussed in \cite{multi-sat}. 

\subsection{Related Works}
Beamforming is an important technique to improve the quality of communication, which can mitigate the intra-beam and inter-beam interference in the multi-beam satellite system by modifying specific radiation patterns of the satellite antenna array and generating directive transmission beams.
In \cite{coexist}, the authors studied the transmit beamforming design for spectral coexistence of satellite and terrestrial networks. 
The authors in \cite{bf1} investigated the cooperative multicast transmission in the integrated terrestrial-satellite network and proposed a cooperative beamforming algorithm to maximize the minimum signal-to-interference-plus-noise ratio (SINR) of users. 

Besides, uniform planar array (UPA) is widely used in satellite communication systems  due to its potential of generating two-dimensional (2D) directive beams. The authors in \cite{UPA-1} and \cite{UPA-2} investigated the downlink transmit design and the application of integrated sensing and communications in LEO satellite communication systems with UPA equipped on the satellite, respectively.
% MIMO
Based on the utilization of large-scale antenna array like UPA, multiple-input multiple-output (MIMO) techniques have been widely studied in satellite communication systems, which enable communication systems to achieve higher flexibility, reliability, and transmission rate.
The authors in \cite{MIMO-sup} and \cite{MIMO-sup2} carried out a review of MIMO-based techniques and proposed the application of the MIMO technology for satellite communications, respectively.
The authors in \cite{MIMO-sup1} studied the MIMO precoding design for multigateway multibeam satellite systems, which can achieve large spectral efficiency (SE) and MIMO transmission design in LEO satellite systems was studied in \cite{MIMO-sup3, MIMO-sup4, MIMO-sup5} and references therein.
Furthermore, massive MIMO communications potentially allow for orders of magnitude improvement in spectral and energy efficiency using relatively simple (linear) processing \cite{MIMO}.
In \cite{bf2}, massive MIMO was first introduced into LEO satellite communication systems and low-complexity statistical channel state information (CSI) based downlink beamforming was proposed to maximize the average signal-to-leakage-plus-noise ratio after Doppler and delay compensations at user terminals.
Later, the authors in \cite{MIMO-sup6} and \cite{MIMO-sup7} investigated the sum rate maximization method and deep learning-based channel prediction in LEO satellite massive MIMO communication system, respectively.

Due to the limitation of the on-board radio-frequency (RF) chains caused by the limited device complexity and transmission power, hybrid analog and digital beamforming is a promising method to balance performances and hardware constraints \cite{hybrid-advantages1, hybrid-advantages2}. 
The authors in \cite{hybrid-ps} proposed an alternate analog-digital beamforming optimization framework for any arbitrary hybrid scheme by using a quadratically constrained quadratic program.
In \cite{hybrid}, a hybrid beamforming strategy was used for massive MIMO LEO satellite communications. 
Later, the authors in \cite{hybrid-RIS} investigated hybrid beamforming method for reconfigurable intelligent surface-assisted secure integrated terrestrial-aerial networks. 

Furthermore, appropriate user scheduling and resource allocation scheme, together with beamforming design, can further improve the communication performance of satellite systems.
In \cite{us3}, a low-complexity CSI based user scheduling algorithm that considers the multigroup multicast nature of the frame-based beamforming system was envisaged.
In \cite{us1}, an adaptive user scheduling method was proposed to mitigate intra-beam and inter-beam interference. 
The authors in \cite{us2} proposed a multilevel clustering algorithm and a cross-cluster grouping algorithm to realize user scheduling. 
Later in \cite{joint}, joint optimization of beamforming design and resource allocation was considered for the terrestrial-satellite cooperation system under the constraints of quality of service (QoS) and backhaul link capacity. 

Besides, full frequency reuse (FFR) is widely adopted \cite{ffr} in satellite systems to improve the SE. However, aggressive frequency reuse will inevitably cause severe inter-beam interference which is also called co-channel interference, and it will affect the multi-satellite cooperative network's performance seriously \cite{cci}. 
As such, multi-satellite collaboration via cooperative digital beamforming is crucial for inter-beam interference mitigation. 
Dirty paper coding (DPC) is an optimal non-linear technique based on known interference at the transmitter which has been proven to reach the same downlink capacity as if there were no interference. However, DPC is not suitable for practical implementation due to its high computational complexity \cite{DPC1, DPC2, DPC3}. 
More practically, linear techniques reduce the complexity at the cost of lower capacity compared with DPC. The authors in \cite{dig-beamf} compared the performances of  several common linear digital beamforming designs: zero-forcing (ZF), regularized ZF and MMSE digital beamformer.

\subsection{Motivations and Contributions}
In previous work, joint optimization was usually performed between beamforming design and resource allocation, while the beamforming design and user scheduling were always separated and independent from each other, where the intrinsic connection between them was not fully exploited. 
To our best knowledge, there is little existing work on the joint design of beamforming and user scheduling for satellite communication networks without the assistance of terrestrial systems. 
Also, single-satellite scenario was often considered in existing works and few of them considered the cooperative transmission of multiple satellites. Although the cooperation of multiple base stations is common in terrestrial communication systems, multi-satellite collaboration in satellite communication systems is still a novel field that needs to be further explored due to the long distance (among satellites and between satellites and users) and severely attenuated transmission signal.
This paper focuses on these points and aims at improving the total SE of the multi-satellite cooperative network by jointly optimizing beamforming and user scheduling. 

In the satellite scenario considered in this work, a series of special issues are different from terrestrial systems and need to be well solved. The motivations and contributions of this paper are summarized as follows:
\begin{itemize}
    \item Restricted by the weight and volume of satellite payloads, the on-board device complexity is limited. In addition, the energy source of the satellite mainly comes from solar energy, so the power for communication transmission is also limited. Thus, the number of on-board RF chains is restricted. Therefore, we propose a hybrid beamforming method based on the hybrid architecture of satellite antennas, whose number of RF chains is far less than the number of antenna elements. The proposed hybrid beamforming method consists of analog beamforming for beam alignment and digital beamforming for interference mitigation.
    \item The distance between a satellite and a GU is much longer than that in terrestrial communication systems. The transmitted signal suffers severe propagation loss, leading to low signal power and achievable rate at the receiver. Thus, the satellite-GU channel cannot be estimated accurately and the CSI feedback need to be simplified as much as possible. Therefore, we propose an analog beamforming method based on codebook, which transmits less content from GUs to satellites and reduces the CSI feedback overhead.
    \item In scenario with multiple GUs and multiple satellites, the links between satellites and GUs should be arranged reasonably to maximize the network total SE. We introduce a discrete variable to indicate the link relation and propose a heuristic user scheduling algorithm with polynomial complexity.
    \item There is intrinsic connection between beamforming and user scheduling. The beamforming result is needed when user scheduling is performing and digital beamforming also needs the information about which GUs are connected with the same satellite. As such, we consider to jointly design the beamforming and user scheduling by means of alternate optimization and propose a joint hybrid beamforming and user scheduling scheme.
    \item Different from most existing works, the proposed algorithms and schemes are applied to not only single-connection scenario, but also multi-connection scenario where one GU can be served by more than one satellite simultaneously to improve the performance.
\end{itemize}

The main findings from simulation results are concluded as follows:
\begin{itemize}
    \item The performance of the proposed joint beamforming and user scheduling scheme can provide remarkable SE improvement compared with its non-joint counterpart under different constellation configurations.
    \item When satellites are serving GUs distributed in a certain latitude range, the constellation inclination and the number of visible satellites are influential to the network's performance. The radiation pattern of the GU antenna will affect the performance differences between the single-connection algorithm and the multi-connection algorithm. Moreover, geographical positions, topological relationships, and the density of GUs bring about performance differences among GUs.
\end{itemize}

\subsection{Paper Organization and Notations}
\textbf{\textit{Organization:}} The rest of this paper is organized as follows. The system architecture, channel model, signal model, and problem formulation are introduced in Section \uppercase\expandafter{\romannumeral2}. 
Subsequently, Section \uppercase\expandafter{\romannumeral3} explains the proposed hybrid beamforming method. The heuristic user scheduling algorithm and two implementation schemes are presented in Section \uppercase\expandafter{\romannumeral4}. 
Section \uppercase\expandafter{\romannumeral5} gives out the simulation results and analysis. Finally we conclude this paper in Section  \uppercase\expandafter{\romannumeral6}.

\textbf{\textit{Notations:}} The following notations are adopted throughout this paper. Boldface upper-case letters refer to matrices and boldface lower-case letters denote column vectors. The transpose, conjugate transpose and inverse are represented by $(\cdot)^{T}, (\cdot)^{H}$, and $(\cdot)^{-1}$, respectively. $\| \cdot \|$ and $\| \cdot \|_F$ stand for the $l_2$-norm of vectors and Frobenius norm of matrices. $(\cdot)^{\dagger}$ denotes the Moore-Penrose pseudo inverse of matrices.
Expectation of a variable is noted by $\mathbb{E}[\cdot]$ and $\otimes$ denotes the Kronecker products of two matrices.

\section{System Model} \label{sec-2}
\subsection{System Architecture} \label{sec-2A}

\begin{figure}[tb]
    \centering
    \includegraphics[width=0.48\textwidth]{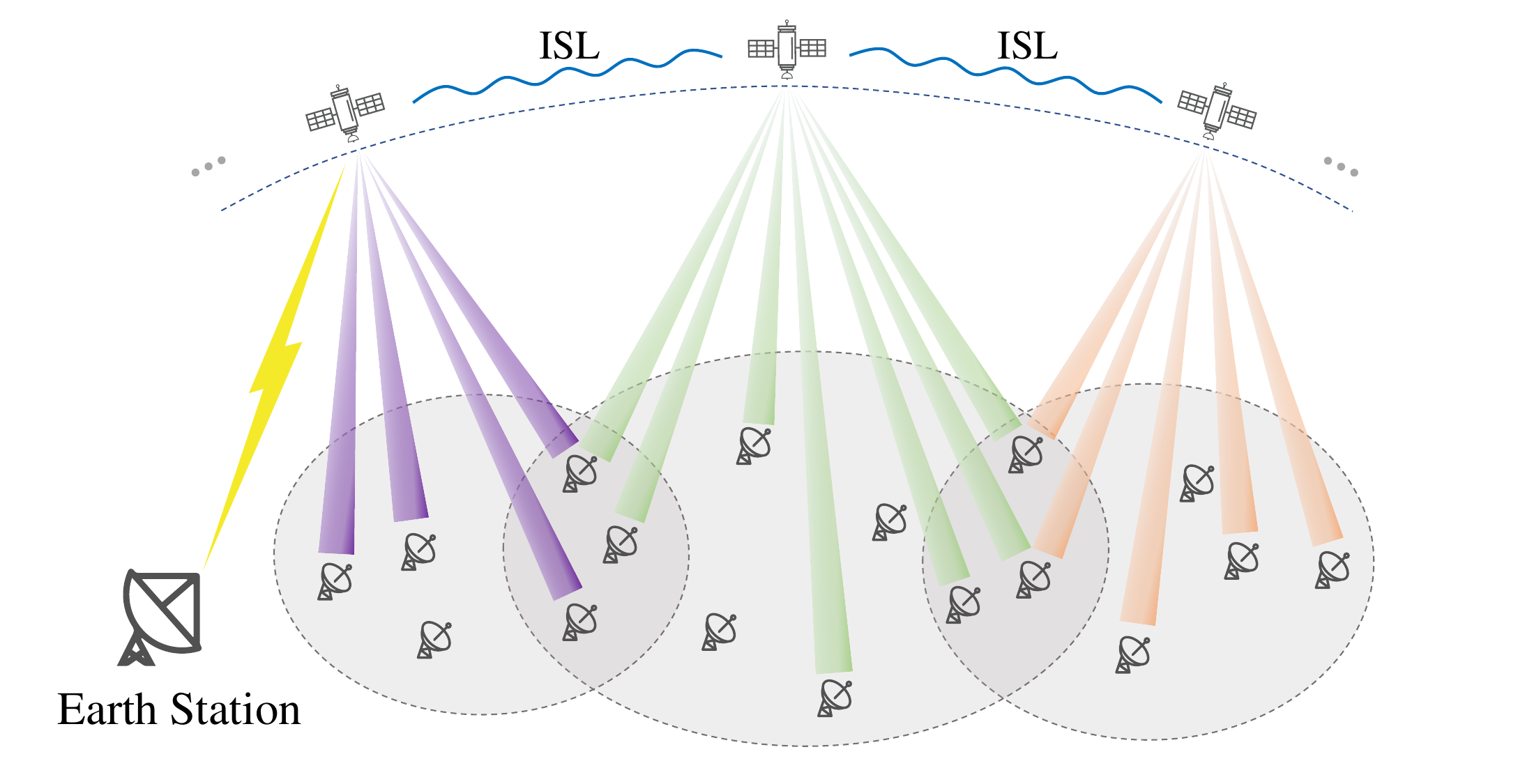}
    \caption{System architecture of the multi-satellite cooperative communication network in this work. ISL denotes inter-satellite link.}
    \label{system}
    % \vspace{-0.7cm}
\end{figure}

As depicted in Fig. \ref{system}, we consider a downlink communication scenario of a multi-satellite communication network in a short time period (millisecond level), where multiple LEO satellites cooperatively provide services for GUs.
We assume that there are $N_u$ GUs requesting services and $N_s$ satellites visible to at least one of these GUs. 
One satellite can serve multiple GUs under its coverage and one GU can also be served by multiple satellites as long as these satellites are visible to the GU.
We use $\mathcal{V}_g$ to denote the set of visible satellites of GU $g$ and $\mathcal{V}_s$ for the set of GUs that satellite $s$ is serving. We also assume that all the satellites are equipped with regenerative payload and belong to an LEO constellation operating in the Ka-band with FFR adopted. Furthermore, there are optical inter-satellite links (ISLs) to exchange data among satellites, and the satellites can perform on-board distributed computing to share computation load \cite{distributed-computing-1, distributed-computing-2}. 
The earth station periodically sends topological relationships of the constellation to the satellites via its line of sight (LoS) and satellites can share the topological relationships through ISLs.

\begin{figure}[tb]
    \centering
    \includegraphics[width=0.48\textwidth]{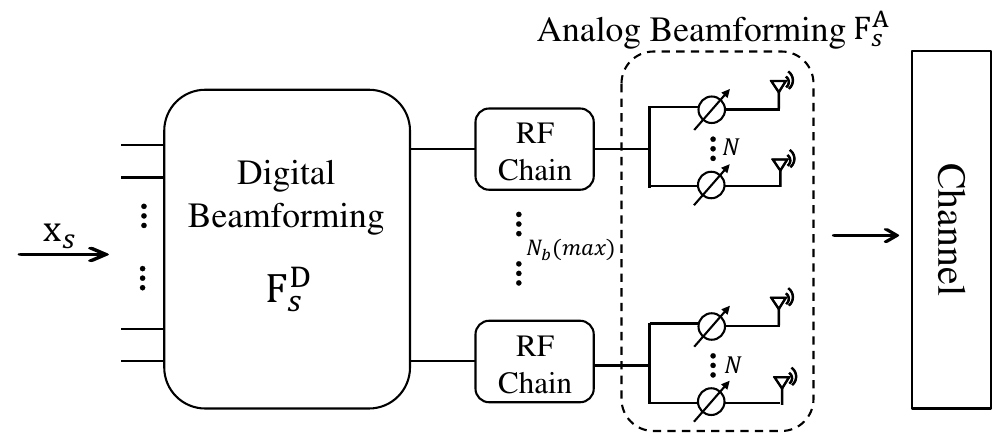}
    \caption{On-board hybrid analog-digital beamforming architecture.}
    \label{hybrid}
    % \vspace{-0.7cm}
\end{figure}

Consider that each GU is equipped with a very small aperture terminal (VSAT) which is a single antenna system, and each satellite is equipped with a UPA facing the Earth. 
Different from purely analog or digital beamforming architecture, the UPA here adopts a hybrid analog-digital beamforming architecture. 
The diagram of the on-board hybrid beamforming architecture is shown in Fig. \ref{hybrid}. The vector $\textbf{x}_s \in \mathbb{C}^{N_u^s \times 1}$ denotes the requested data vector of GUs that satellite $s$ is serving, where $N_u^s$ is the number of GUs that satellite $s$ is serving. $\mathbf{F}_s^\text{A} \in \mathbb{C}^{N \times N_u^s}$ and $\mathbf{F}_s^\text{D} \in \mathbb{C}^{N_u^s \times N_u^s}$ are the analog beamforming matrix and the digital beamforming matrix of satellite $s$, respectively, and they will be further discussed in the following sections. 
The UPA is composed of $N_b = N_{x}^\text{sub} \times N_{y}^\text{sub}$ sub-arrays where $x$ denotes the axis pointing in the direction of the satellite's movement and $y$ denotes the axis pointing in the direction orthogonal to the satellite's movement. 
Each sub-array consists of $N = N_x \times N_y$ antenna elements and is connected with one RF chain, generating one independent spot beam. 
One satellite can simultaneously generate $N_b$ independent spot beams at most, and each independent beam serves one single GU at any given moment \cite{Channel-Model}, which indicates that one satellite can provide services for up to $N_b$ GUs simultaneously.

\subsection{Channel Model}
\begin{figure}[tb]
    \centering
    \includegraphics[width=0.48\textwidth]{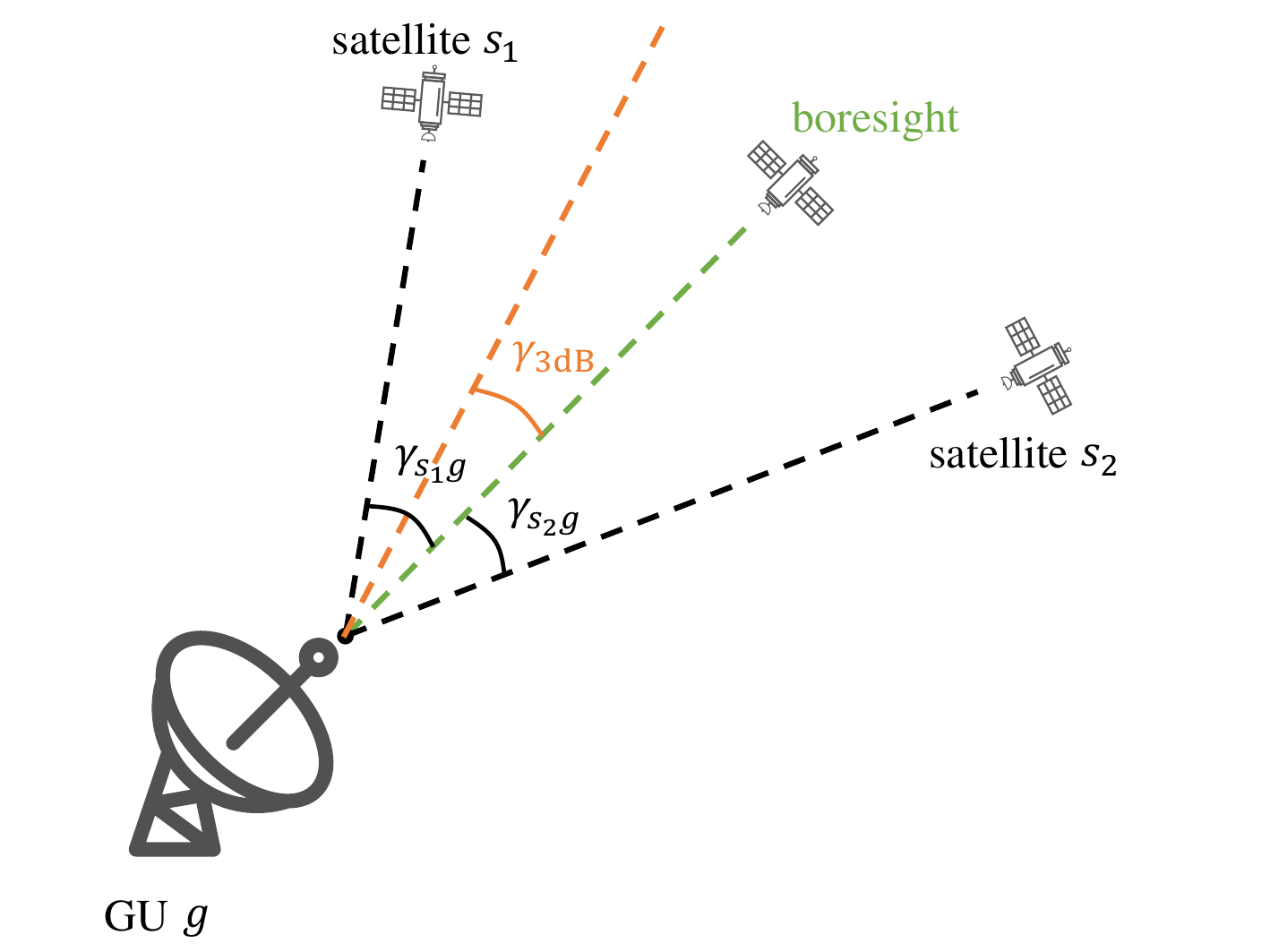}
    \caption{Diagram of GU antenna beam and angles.} %$\gamma_{sg}$ and $\gamma_{\text{3dB}}$
    \label{gain}
    % \vspace{-0.5cm}
\end{figure}

The propagation channel in our study is modeled according to the technical reports of the 3rd Generation Partnership Project (3GPP) and the International Telecommunication Union Radiocommunication Sector (ITU-R) \cite{3GPP-38811, ITU-R-676, ITU-R-681, 3GPP-38901}.
We consider the scenario where the satellite-GU link suffers from no rain and cloud attenuation, and all the GUs are distributed in suburban areas.
The multiple-input single-output channel between satellite $s$, for $s \in \{1,2,\dots,N_s\}$ and GU $g$ for $g \in \{1,2,\dots,N_u\}$ can be modeled as
\begin{equation}  \label{eq-channel}
    \textbf{h}_{sg} = \xi_{sg} \cdot \textbf{h}_{sg_{s}},
\end{equation}
where $\xi_{sg}$ and $\textbf{h}_{sg_{s}} \in \mathbb{C}^{N \times 1}$ stand for the radio propagation loss and the small-scale fading channel which follows a Loo distribution, respectively. Here, $\xi_{sg}$ includes the effects of the large-scale path loss (PL) and the antenna gains, which can be expressed as
\begin{equation}
    \xi_{sg} = \sqrt{G_\text{S} G_{\text{GU},sg}} \cdot 10^{-\frac{1}{10}\text{PL[dB]}},
\end{equation}
where $G_\text{S}$ is the gain of the satellite UPA and $G_{\text{GU},sg}$ is the GU-side antenna gain between satellite $s$ and GU $g$. We define $\gamma_{sg}$ as the off-boresight angle of the LoS link between satellite $s$ and GU $g$, and $\gamma_{\text{3dB}}$ as the 3-dB angle of the GU antenna, as shown in Fig. \ref{gain}. Thus, the GU-side antenna gain can be approximated by \cite{gain}
\begin{equation} \label{eq-gain}
    G_{\text{GU},sg} \footnote{The antenna model in Eq. (\ref{eq-gain}) differs from that in our scenario, but Eq. (3) can be utilized to approximate the gain difference of various off-boresight angles without loss of generality.} \approx G_{\text{max}} \left( \frac{J_1(u_{sg})}{2u_{sg}} + 36\frac{J_3(u_{sg})}{u_{sg}^3} \right)^2,
\end{equation}
where $G_{\text{max}}$ is the maximum gain at the boresight, $u_{sg} = 2.07123 \sin{\gamma_{sg}}/\sin{\gamma_{\text{3dB}}}$, $J_1(\cdot)$ and $J_3(\cdot)$ denote the first-kind Bessel function of order 1 and order 3, respectively. The GU antenna gain $G_{\text{GU},sg}$ will approach $G_\text{max}$ when $\gamma_{sg} \to 0$ and approach $\frac{1}{2} G_\text{max}$ when $\gamma_{sg} \to \gamma_{\text{3dB}}$. When a GU is served by multiple satellites simultaneously, only the satellite at the boresight can obtain the maximum GU-side antenna gain and the others will suffer a gain reduction. 

The large-scale PL can be calculated as follows \cite{3GPP-38811}:
\begin{equation}
    \text{PL}[\text{dB}] = \text{PL}_\text{b}[\text{dB}] + \text{PL}_\text{g}[\text{dB}] + \text{PL}_\text{s}[\text{dB}],
\end{equation}
where $\text{PL}_\text{b}$ denotes the basic path loss, $\text{PL}_\text{g}$ denotes the attenuation due to atmospheric gasses, and $\text{PL}_\text{s}$ denotes the attenuation due to either ionospheric or tropospheric scintillation. 
The basic path loss can be expressed as \cite{3GPP-38811}
\begin{equation}
    \text{PL}_\text{b}[\text{dB}] = \text{FSPL}(d_0, f_c)[\text{dB}] + \text{SF}[\text{dB}] + \text{CL}[\text{dB}],
\end{equation}
where FSPL denotes the free space path loss depending on the transmission distance $d_0$ and the carrier frequency $f_c$, SF denotes shadow fading modeled by a log-normal distribution, and CL denotes clutter loss which is negligible and should be set to 0 dB because of the assumption of LoS condition.
The path loss due to atmospheric gasses $\text{PL}_\text{g}$ depends mainly on frequency, elevation angle, altitude above sea level, and water vapour density (absolute humidity), which is modeled according to \cite{ITU-R-676}. The path loss due to scintillation $\text{PL}_\text{s}$ contains tropospheric scintillation only because ionospheric scintillation can be neglected when the satellites are working in the Ka-band.

The small-scale fading channel model follows a Loo distribution of the GOOD state in \cite{ITU-R-681}, where the received signal is the sum of two components: the direct path and the diffuse multipath. We assume that there are $N_\text{cl}$ clusters with $N_\text{ray}$ propagation paths in each cluster. The small-scale fading channel $\textbf{h}_{sg_s}$ can be modeled as
\begin{equation}
    \begin{split}
        \textbf{h}_{sg_s} = \delta & \bigg( m_0 \, \textbf{a}_\text{T}(\phi_{sg}, \theta_{sg}) 
         + \sum_{l=1}^{N_\text{cl}} \sum_{i=1}^{N_\text{ray}} m_{li} \textbf{a}_\text{T}(\phi_{sg,li}, \theta_{sg,li}) \bigg),
    \end{split}
\end{equation}
where $m_0$ and $m_{li}$ are complex coefficients of the direct path and the diffuse multipath, respectively. The amplitude of $m_0$ is subject to the normal distribution, while the amplitude of $m_{li}$ obeys the Rayleigh distribution \cite{ITU-R-681}. The phases of both $m_0$ and $m_{li}$ follow a uniform distribution from 0 to 2$\pi$. 
% The specific parameters are given as in \cite{ITU-R-681}. 
The normalization factor $\delta$ is introduced to satisfy $\mathbb{E} \left[ \| \textbf{h}_{sg_s} \|^{2} \right] = 1$. 
$\phi_{sg}$ and $\theta_{sg}$ denote the azimuth and elevation angles of the direct path from the perspective of satellite. While $\phi_{sg,li}$ and $\theta_{sg,li}$ are the azimuth and elevation angles of the $i$-th ray in the $l$-th cluster which belongs to the diffuse multipath, and they can be obtained using the method in \cite{3GPP-38811, 3GPP-38901}.
The vector $\textbf{a}_\text{T}(\phi, \theta) \in \mathbb{C}^{N \times 1}$ is the normalized antenna steering vector of the satellite's sub-array, which can be written as

\begin{equation}
    \begin{aligned}
        \textbf{a}_\text{T}(\phi, \theta) = \frac{1}{\sqrt{N}} \left[ 1, \dots, e^{-j \frac{2\pi}{\lambda} d \left( p \cos{\theta} \cos{\phi} + q \cos{\theta} \sin{\phi} \right)}, \dots, \right.\\
        \left. e^{-j \frac{2\pi}{\lambda} d \left( \left( N_x - 1 \right) \cos{\theta} \cos{\phi} + \left( N_y - 1 \right) \cos{\theta} \sin{\phi} \right) } \right]^T,
    \end{aligned}
\end{equation}
where $\lambda$ and $d$ are the carrier wavelength and the antenna element spacing, respectively. In our study, we assume $d = \frac{\lambda}{2}$ to guarantee that there is no grating lobe when a beam is steered towards $\pm\, 90^{\circ}$. And $p$, $q$ are the antenna element indices which are integers satisfying $p \in [0, N_x - 1]$, $q \in [0, N_y - 1]$. 

Besides, there is significant Doppler effect due to the motion of the satellite. Nevertheless, the satellite's movement is highly predictable and the Doppler compensation has been investigated in many existing works. The relevant technique is mature so that we will not consider it in our work.

\subsection{Signal Model and Problem Formulation} \label{sec-2C}
As mentioned in Section \ref{sec-2A}, multiple satellites provide communication services for their GUs cooperatively with FFR. Therefore, each GU will suffer interference from all the satellites that are in this GU's LoS. 
When receiving signals, the GU antenna will aim at the first serving satellite (to be detailed later on), and the intended signal and interference from this satellite will experience the maximum gain of the GU antenna. 
While the interference and intended signals from other visible satellites will experience a gain reduction due to the off-boresight angles of these satellite-GU links and the narrow-beam characteristics of the VSAT considered herein.
Note that the set of visible satellites of each GU is known, but the specific links between satellites and GUs are unknown and need to be determined. Based on this fact, we introduce a discrete variable $\alpha_{sg}$ to indicate whether a link is established, where $\alpha_{sg} = 1$ if satellite $s$ is providing service for GU $g$, and $\alpha_{sg} = 0$ otherwise.

The received signal of GU $g$ can be expressed as
\begin{equation} \label{eq-signalmodel}
    y_{g} = \sum_{s \in \mathcal{V}_g} \alpha_{sg} \textbf{h}_{sg}^\text{H} \textbf{w}_{sg} x_{g} + \sum_{g^{'} \neq \, g} \sum_{s \in \mathcal{V}_g} \alpha_{sg^{'}} \textbf{h}_{sg}^\text{H} \textbf{w}_{sg^{'}} x_{g^{'}} + n_g,
\end{equation}
where $\textbf{h}_{sg} \in \mathbb{C}^{N \times 1}$ is the channel vector between satellite $s$ and GU $g$, $\textbf{w}_{sg} \in \mathbb{C}^{N \times 1}$ represents the beamforming vector, and $x_g$ is the requested data of GU $g$, which is assumed to be independent and satisfy $\mathbb{E} \left( |x_g|^2 \right) = 1$. 
The first term in (\ref{eq-signalmodel}) is the intended data for GU $g$, the second term is the interference coming from the communication services for the other GUs, and the third term is the complex additive white Gaussian noise following the distribution $\mathcal{CN} \left(0, \sigma_s^2 \right)$, where $\sigma_s^2$ denotes the noise power.

Then the received SINR of GU $g$ can be obtained as
\begin{equation}
    \gamma_{g} = \frac{|\sum_{s \in \mathcal{V}_g} \alpha_{sg} \textbf{h}_{sg}^\text{H} \textbf{w}_{sg}|^{2}}
    {\sum_{g^{'} \not = g} | \sum_{s \in \mathcal{V}_g} \alpha_{sg^{'}} \textbf{h}_{sg}^\text{H} \textbf{w}_{sg^{'}}|^{2} + \sigma_{s}^{2}}.
\end{equation}

According to the Shannon theorem, the SE per channel use of GU $g$ can be calculated by
\begin{equation}
    R_{g} = \log_{2}{(1 + \gamma_{g})}.
\end{equation}
As a result, the total SE of the multi-satellite cooperative network can be expressed as
\begin{equation} \label{total-SE}
    R = \sum_{g=1}^{N_u} R_{g}.
\end{equation}

Our objective is to maximize the total SE of the multi-satellite cooperative network by means of calculating the beamforming vector $\textbf{w}_{sg}$ and adjusting the satellite-GU links. Therefore, we formulate the optimization problem $\left(\mathcal{OP} \right)$ as follows:
\begin{align}
    \mathcal{OP}: \quad &
    \max_{\{\textbf{w}_{sg},\alpha_{sg} \}} \qquad \sum_{g=1}^{N_u} \log_{2}{(1 + \gamma_{g})}\\
    s.t.\quad & C_1:\sum_{g} \alpha_{sg} \| \textbf{w}_{sg} \|^{2} \leq P_\text{T}, \, \forall s \in \{1,2,\dots,N_s\},\\
    & C_2:\sum_{g} \alpha_{sg} \leq N_{b}, \, \forall s \in \{1,2,\dots,N_s\},\\
    & C_3:\sum_{s \in \mathcal{V}_g} \alpha_{sg} \geq 1, \, \forall g \in \{1,2,\dots,N_u\},\\
    & C_4:\alpha_{sg} \in \{ 0,1 \}, \, \forall s,g.
\end{align}
Here, constraint $C_1$ is the power constraint of each satellite where the total power of each satellite can not exceed $P_\text{T}$. 
Constraint $C_2$ means that the number of GUs connected to the same satellite cannot exceed the maximum number of spot beams that one satellite can generate, namely $N_b$. 
Constraint $C_3$ is the GU connection constraint. 
Notably, the $\mathcal{OP}$ is not always feasible because of the coexistence of $C_2$ and $C_3$. When $C_2$ and $C_3$ contradict each other, we give priority to guaranteeing $C_2$ and try to connect as many GUs as possible.

The $\mathcal{OP}$ cannot be solved directly due to the fact that the objective function is non-convex and the beamforming vectors $\textbf{w}_{sg}$ are coupled with the link indicators $\alpha_{sg}$ in the objective function and constraint $C_1$. Thus, we propose to solve it by the following steps:
\begin{enumerate}
    \item Given $\textbf{h}_{sg}$ in (\ref{eq-channel}), the analog beamforming vector $\textbf{w}_{sg}^\text{A}$ is obtained based on a discrete Fourier transform (DFT) codebook (with more details to be shown in Algorithm \ref{Codebook-Ana}).
    \item The vector $\textbf{w}_{sg}^\text{A}$ is set as the initial beamforming vector.
    \item Based on 2), we propose two schemes to solve $\mathcal{OP}$, which will be further discussed in the following sections.
\end{enumerate}

\section{Hybrid Beamforming} \label{sec-3}
\subsection{Analog Beamforming Based on Codebook} \label{sec-3A}
We consider analog beamforming for beam alignment based on a codebook, which is widely used in terrestrial systems. Codebook-based beamforming design can reduce the overhead of CSI feedback. In this work, we assume that the GU side has perfect CSI and the GU-satellite CSI feedback link is lossless.
As described in Section \ref{sec-2A}, each satellite is equipped with a UPA, thus the 2D DFT codebook is applicable and it can be seen as the synthesis of two 1D DFT codebooks in the directions of x and y axes, $\mathbf{D}_x,\mathbf{D}_y$. 

When $d = \frac{\lambda}{2}$, the 1D DFT codebook in the x-axis $\mathbf{D}_x$ can be written as
\begin{equation}
    \mathbf{D}_x = \frac{1}{\sqrt{N_{x}}}
        \begin{bmatrix}
        1     & \cdots  &  1   &  \cdots\\ 
        e^{j\pi\sin{\phi_0^x}}  &  \cdots  &  e^{j\pi\sin{\phi_{k}^x}}  &  \cdots\\
        e^{j\pi2\sin{\phi_0^x}}   &  \cdots  &  e^{j\pi2\sin{\phi_{k}^x}}  &  \cdots\\
        \vdots & \vdots      & \vdots  &  \cdots\\
        e^{j\pi(N_x-1)\sin{\phi_0^x}}   & \cdots & e^{j\pi(N_x-1)\sin{\phi_{k}^x}}  &  \cdots
        \end{bmatrix},
\end{equation}
where $\sin{\phi_k^x} = 1-\frac{2}{N_x}k,\,k = 0,1,\cdots,N_x-1$.
Then the $N \times N$ 2D DFT codebook matrix \cite{DFT-Codebook}, denoted as $\mathbf{D}$, can be expressed  as
\begin{equation}
    \mathbf{D} = \mathbf{D}_x \otimes \mathbf{D}_y,
\end{equation}
where the 2D DFT codebook $\mathbf{D}$ consists of $N$ independent codewords which are orthogonal to each other and each codeword corresponds to a directive beam towards the earth.

\begin{algorithm}[tb] 
    % \SetAlgoLined %显示end
    \caption{Codebook-Based Analog Beamforming}
    \label{Codebook-Ana}
    \KwIn{GU-side channel vector $\textbf{h}_{sg},\forall s,g$, codebook $\mathbf{D}$.}
    \KwOut{Analog beamforming vector $\textbf{w}_{sg}^\text{A}$.}
    For each codeword $\mathbf{D}_{:,k}$, calculate $|\textbf{h}_{sg}^\text{H} \mathbf{D}_{:,k}|^2$ and find the best $K$ codewords maximizing it, $\textbf{c}_1,\dots,\textbf{c}_K$\;
    $\mathbf{D}_K = [\textbf{c}_1,\dots,\textbf{c}_K]$, solve the equations $\mathbf{D}_K \textbf{x} = \textbf{h}_{sg}$ and obtain the least square solution $\hat{\textbf{x}} = (\mathbf{D}_K)^{\dagger} \textbf{h}_{sg}$\;
    The GU sends the codeword combination coefficients $\hat{\textbf{x}}$ and codewords' indices to the satellite\;
    Satellite-side combination: $\textbf{w}_{sg}^{'} = \mathbf{D}_K \hat{\textbf{x}}$\;
    \For{$i \in [1, N]$}{$\textbf{w}_{sg}^\text{A}(i) = \textbf{w}_{sg}^{'}(i) \frac{\frac{1}{\sqrt{N}}}{|\textbf{w}_{sg}^{'}(i)|}$}
\end{algorithm}

The core idea of analog beamforming herein is to select the best $K$ ($\leq N$) codewords from $\mathbf{D}$ and combine them into a new codeword that satisfies the equal-amplitude constraint of analog beamforming. 
Considering the communication overhead of CSI feedback, $K$ should not be too large and we assume $K = 4$ based on the results of trial simulations. The steps of the proposed analog beamforming method are given in Algorithm \ref{Codebook-Ana}. 
Notably, the trajectory of satellites is predictable according to the prior information in terms of the orbit and movement of satellites so that the GU can track the satellite of interest in real time and send it messages. 
To begin with, the best $K$ codewords are selected based on the GU-side perfect CSI and the codeword combination coefficients are calculated on the GU side. Then the GU sends the codewords' indices and codeword combination coefficients to the satellite. 
Based on these, the satellite linearly combines the $K$ codewords in accordance to the coefficients and adjusts each component of the beamforming vector to satisfy the equal-amplitude constraint of analog beamforming.

\subsection{Digital Beamforming}
Based on the link information between satellites and GUs, the channel matrix of satellite $s$ can be written as
\begin{equation} \label{eq-channelmat}
    \mathbf{H}_s = \left[\dots, \textbf{h}_{sg}, \dots \right]^\text{H} \in \mathbb{C}^{N_u^s \times N}, \, g \in \mathcal{V}_s.
\end{equation}
% where $N_u^s$ is the number of GUs that satellite $s$ is serving. 
Similarly, the analog beamforming matrix of satellite $s$ can be written as
\begin{equation} \label{eq-anabeamf}
    \mathbf{F}_s^\text{A} = \left[\dots, \textbf{w}_{sg}^\text{A}, \dots \right] \in \mathbb{C}^{N \times N_u^s}, \, g \in \mathcal{V}_s.
\end{equation}
Hence, we can write the generalized channel matrix between satellite $s$ and the GUs as
\begin{equation} \label{eq-generalized-channel}
    \widetilde{\mathbf{H}}_s = \mathbf{H}_s \mathbf{F}_s^\text{A},
\end{equation}
and the channel matrix in Eq. (\ref{eq-generalized-channel}) can be obtained according to the CSI feedback as stated in Section \ref{sec-3A}.
Thus the hybrid beamforming is reduced to a digital beamforming problem to mitigate the inter-beam interference of satellite $s$. In this work, we adopt the regularized ZF and the corresponding digital beamforming matrix is
\begin{equation} \label{eq-digbeamf}
    \mathbf{F}_s^\text{D} = \sqrt{\eta} \, \widetilde{\mathbf{H}}_s^\text{H} (\widetilde{\mathbf{H}}_s \widetilde{\mathbf{H}}_s^\text{H} + \beta \mathbf{I}_{N_u^s})^{-1} \in \mathbb{C}^{N_u^s \times N_u^s},
\end{equation}
where $\sqrt{\eta}$ is a power scaling factor to guarantee the satellite is operating at its maximum power, and $\beta$ is an adjustable parameter where $\beta_\text{opt} = \frac{N_u^s \sigma_s^2}{P_\text{T}}$ in the large system limit \cite{ChannelInversion1, ChannelInversion2}.

Thus, combining (\ref{eq-anabeamf}) and (\ref{eq-digbeamf}), the hybrid beamforming matrix can be expressed as
\begin{equation} \label{eq-hybeamf}
    \mathbf{F}_s^\text{HY} = \mathbf{F}_s^\text{A} \cdot \mathbf{F}_s^\text{D} = \left[\dots, \textbf{w}_{sg}, \dots \right] \in \mathbb{C}^{N \times N_u^s}, \, g \in \mathcal{V}_s.
\end{equation}

\begin{algorithm}[tb]
    % \SetAlgoLined %显示end
    \caption{Single-Connection Heuristic User Scheduling Algorithm}
    \label{PUS}
    \KwIn{Channel matrix $\mathbf{H}_s$, beamforming matrix $\mathbf{F}_s$, and the set of visible satellites $\mathcal{V}_g$, $\forall s \in \{1,2,\dots,N_s\}$, $\forall g \in \{1,2,\dots,N_u\}$.}
    \KwOut{Single-connection link matrix $\mathbf{L}$, set $\mathcal{S}$.}
    Initialize $\mathcal{S}=\{1,\dots,N_s\}, \mathcal{G}_n \big|_{n = N_u}=\{1,\dots,N_u\}, \mathbf{L}=\mathbf{0}$\;
    % \tcp{procedure SINGLE CONNECTION}
    \For{$g \in [1, N_u]$}{
        \If{\rm length($\mathcal{V}_g$) $== 1$}{
            $\mathbf{L}(\mathcal{V}_g,g) = 1$, remove $g$ from $\mathcal{G}_n$ and $n=n-1$\;
        }
    }
    \Repeat{$n==0$}{
        \For{{\rm each possible link} $\mathbf{L}_{sg},\, s \in \mathcal{S}, g \in \mathcal{G}_n$}{
        $\triangle R_{sg} = \mathcal{R}(\mathbf{L}+\mathbf{L}_{sg}, \mathbf{H}_s, \mathbf{F}_s) - \mathcal{R}(\mathbf{L}, \mathbf{H}_s, \mathbf{F}_s)$\;
        }
        $[\hat{s},\hat{g}] = \mathop{\arg\max}\limits_{s,g} \triangle R_{sg}$\;
        \eIf{{\rm satellite} $\hat{s}$ {\rm has spare resource}}{
            $\mathbf{L}(\hat{s},\hat{g}) = 1$, remove $\hat{g}$ from $\mathcal{G}_n$ and $n=n-1$\;
        }{
            remove $\hat{s}$ from $\mathcal{S}$\;
        }
    }
\end{algorithm}

\section{User Scheduling and Implementation Schemes}
\subsection{User Scheduling}
As described in Section \ref{sec-2C}, the links between multiple satellites and GUs need to be determined. The optimal exhaustive search is infeasible here since the computational complexity grows exponentially with the number of GUs. Thus, we propose a heuristic user scheduling algorithm which can achieve a good performance with polynomial complexity.
In existing researches, the single-connection condition is usually assumed where each GU connects with only one satellite. The single-connection user scheduling algorithm is shown in Algorithm \ref{PUS}. 
Therein, $\mathbf{L}$ denotes an $N_s \times N_u$ link matrix where $\alpha_{sg}$ lies in its $s$-th row and $g$-th column, and $\mathbf{L}_{sg}$ denotes the matrix whose $s$-th row and $g$-th column is 1 and other places are 0. The set of satellites that have spare resource is denoted as $\mathcal{S}$. Here, the resource constraint is that the number of GUs that one satellite is serving simultaneously can not exceed $N_b$. The set of unserved GUs is denoted by $\mathcal{G}_n$, where $n$ is the number of GUs in this set. $\triangle R$ indicates the increment of total SE. The total SE is mainly associated with the link matrix $\mathbf{L}$, channel matrix $\mathbf{H}_s$ and the beamforming matrix $\mathbf{F}_s$, which is equivalent to (\ref{total-SE}) and can be abstracted as
\begin{equation} \label{eq-R}
    R = \mathcal{R}(\mathbf{L}, \mathbf{H}_s, \mathbf{F}_s),
\end{equation}
where $\mathcal{R}$ indicates a function for calculating the total SE.
The single-connection algorithm follows the steps below:
\begin{enumerate}
    \item Find the GUs who can only see one satellite and establish their connections first.
    \item Calculate the increment of network's total SE when establishing each possible link.
    \item Find the largest total SE increment among all the possible links and record the link.
    \item Check whether the resource constraint is satisfied: if the link meets the resource constraint, establish the connection; otherwise, remove it and find the next best link until the resource constraint is met.
\end{enumerate}

\begin{algorithm}[tb]
    % \SetAlgoLined %显示end
    \caption{Multi-Connection Heuristic User Scheduling Algorithm}
    \label{PUS-Multiple}
    \KwIn{Channel matrix $\mathbf{H}_s$, beamforming matrix $\mathbf{F}_s$, single-connection link matrix $\mathbf{L}$, output set $\mathcal{S}$ of Algorithm \ref{PUS}, and the set of visible satellites $\mathcal{V}_g$, $\forall s \in \{1,2,\dots,N_s\}$, $\forall g \in \{1,2,\dots,N_u\}$.}
    \KwOut{Multi-connection link matrix $\mathbf{L}$.}
    \Repeat{\rm $\triangle R_{\hat{s}\hat{g}} \leq 0$ or length($\mathcal{S}$) $ == 0$}{
        \For{\rm each possible link $\mathbf{L}_{sg}, \, s \in \mathcal{S}$}{
            $\triangle R_{sg} = \mathcal{R}(\mathbf{L}+\mathbf{L}_{sg}, \mathbf{H}_s, \mathbf{F}_s) - \mathcal{R}(\mathbf{L}, \mathbf{H}_s, \mathbf{F}_s)$\;
        }
        $[\hat{s},\hat{g}] = \mathop{\arg\max}\limits_{s,g} \triangle R_{sg}$\;
        \If{$\triangle R_{\hat{s}\hat{g}} > 0$}{
            \eIf{\rm satellite $\hat{s}$ has spare resource}{
                $\mathbf{L}(\hat{s},\hat{g}) = 1$\;
            }{
                remove $\hat{s}$ from $\mathcal{S}$\;
            }
        }
    }
\end{algorithm}

After the single-connection links between satellites and GUs are determined, we establish multi-connection links to improve the total SE of the network. 
As described in Section \ref{sec-2C}, when receiving signals, the main lobe of the GU antenna points at the serving satellite determined through Algorithm \ref{PUS}, thus signals from other satellites will suffer an antenna gain reduction. The different receiving gains of different GU-satellite links must be taken into consideration. 
The multi-connection user scheduling algorithm based on the result of single-connection algorithm is shown in Algorithm \ref{PUS-Multiple}, which follows similar steps to 2), 3), 4) above, but the stopping criterion is different.

\begin{algorithm}[tbp]
    % \SetAlgoLined %显示end
    \caption{JHU Scheme}
    \label{JHU}
    \KwIn{Channel matrix $\mathbf{H}_s$, beamforming matrix $\mathbf{F}_s$, and the set of visible satellites $\mathcal{V}_g$, $\forall s \in \{1,2,\dots,N_s\}$, $\forall g \in \{1,2,\dots,N_u\}$.}
    \KwOut{Multi-connection link matrix $\mathbf{L}$, set $\mathcal{S}$.}
    Initialize $\mathcal{S}=\{1,\dots,N_s\}, \mathcal{G}_n \big|_{n = N_u}=\{1,\dots,N_u\}, \mathbf{L}=\mathbf{0}$\;
    \tcp{SINGLE-CONNECTION procedure}
    \For{$g \in [1, N_u]$}{
        \If{\rm length($\mathcal{V}_g$) $== 1$}{
            $\mathbf{L}(\mathcal{V}_g,g) = 1$,
            remove $g$ from $\mathcal{G}_n$ and $n=n-1$\;
        }
    }
    \Repeat{$n==0$}{
        \For{{\rm each possible link} $\mathbf{L}_{sg},\, s \in \mathcal{S}, g \in \mathcal{G}_n$}{
        $\triangle R_{sg} =\, \mathcal{R}(\mathbf{L}+\mathbf{L}_{sg}, \mathbf{H}_s, \mathcal{F}(\mathbf{L}+\mathbf{L}_{sg}, \mathbf{H}_s, \mathbf{F}_s^\text{A}))
        - \mathcal{R}(\mathbf{L}, \mathbf{H}_s, \mathcal{F}(\mathbf{L}, \mathbf{H}_s, \mathbf{F}_s^\text{A}))$\;
        }
        $[\hat{s},\hat{g}] = \mathop{\arg\max}\limits_{s,g} \triangle R_{sg}$\;
        \eIf{{\rm satellite} $\hat{s}$ {\rm has spare resource}}{
            $\mathbf{L}(\hat{s},\hat{g}) = 1$, remove $\hat{g}$ from $\mathcal{G}_n$ and $n=n-1$\;
        }{
            remove $\hat{s}$ from $\mathcal{S}$\;
        }
    }
    \tcp{MULTI-CONNECTION procedure}
    \Repeat{\rm $\triangle R_{\hat{s}\hat{g}} \leq 0$ or length($\mathcal{S}$) $ == 0$}{
        \For{\rm each possible link $\mathbf{L}_{sg}, \, s \in \mathcal{S}$}{
            $\triangle R_{sg} =\, \mathcal{R}(\mathbf{L}+\mathbf{L}_{sg}, \mathbf{H}_s, \mathcal{F}(\mathbf{L}+\mathbf{L}_{sg}, \mathbf{H}_s, \mathbf{F}_s^\text{A}))
        - \mathcal{R}(\mathbf{L}, \mathbf{H}_s, \mathcal{F}(\mathbf{L}, \mathbf{H}_s, \mathbf{F}_s^\text{A}))$\;
        }
        $[\hat{s},\hat{g}] = \mathop{\arg\max}\limits_{s,g} \triangle R_{sg}$\;
        \If{$\triangle R_{\hat{s}\hat{g}} > 0$}{
            \eIf{\rm satellite $\hat{s}$ has spare resource}{
                $\mathbf{L}(\hat{s},\hat{g}) = 1$\;
            }{
                remove $\hat{s}$ from $\mathcal{S}$\;
            }
        }
    }
\end{algorithm}

\subsection{Separate \& Joint Hybrid Beamforming and User Scheduling Schemes}
In Section \ref{sec-3}, we have introduced the hybrid beamforming method. Within the hybrid beamforming, the analog beamforming can be completed independently. As for the digital beamforming and user scheduling, there are two ways:
\subsubsection{\textbf{\textit{Separate (SHU)}}}
In the SHU scheme, we perform digital beamforming and user scheduling separately and independently. Analog beamforming matrix $\mathbf{F}_s^\text{A}$ is taken as the input of Algorithm \ref{PUS} and user scheduling is performed based on $\mathbf{F}_s^\text{A}$, i.e., $\mathbf{F}_s = \mathbf{F}_s^\text{A}$. Digital beamforming is conducted after user scheduling according to (\ref{eq-digbeamf}), utilizing the final links to mitigate the interference and improve the performance. Finally, we use $\mathbf{F}_s^\text{HY}$ in (\ref{eq-hybeamf}) to calculate the total SE when SHU is adopted.
\subsubsection{\textbf{\textit{Joint (JHU)}}}
The JHU scheme is based on alternating optimization. Different from SHU, user scheduling and digital beamforming in JHU are designed jointly. The beamforming matrix is updated in real time within the user scheduling. As in SHU, $\mathbf{F}_s^\text{A}$ is taken as the initial input of Algorithm \ref{PUS}. The difference is that each time before calculating the SE increment $\triangle R_{sg}$, the hybrid beamforming matrix $\mathbf{F}_s^\text{HY}$ is computed based on the current link matrix. We abstract (\ref{eq-digbeamf}) and (\ref{eq-hybeamf}) as
\begin{equation} \label{eq-hy}
    \mathbf{F}_s^\text{HY} = \mathcal{F}(\mathbf{L}, \mathbf{H}_s, \mathbf{F}_s^\text{A}),
\end{equation}
where $\mathcal{F}$ is a function for calculating the hybrid beamforming matrix. The algorithm flow of JHU scheme is shown in Algorithm \ref{JHU}.
The computational complexity of the proposed algorithms and exhaustive search is presented in Table \ref{complexity}. $M_\text{R}$ denotes the number of calculation for the total SE according to (\ref{eq-R}). $M_\text{HY}$ denotes the number of calculation for hybrid beamforming matrix according to (\ref{eq-hy}) and one calculation of hybrid beamforming matrix contains one calculation of matrix addition, one calculation of matrix inversion, and four calculations of matrix multiplication.

% non-blue
\begin{table}[bt]
    \centering
    \caption{Computational Complexity Analysis}
    \label{complexity}
    \begin{tabular}{c|c|c|c}
        \hline
         & Exhaustive Search & SHU & JHU\\
        \hline
        $M_\text{R}$ & $\mathcal{O}(2^{N_s N_u})$ & $\mathcal{O}(N_u^2 N_s)$ &  $\mathcal{O}(N_u^2 N_s)$\\
        $M_\text{HY}$ & $\mathcal{O}(2^{N_s N_u} N_s)$ & $N_s$ & $\mathcal{O}(N_u^2 N_s)$\\
        \hline
    \end{tabular}
    % \vspace{-0.5cm}
\end{table}

\section{Performance Evaluation}
In this section, the azimuth angle, elevation angle, and range data between satellites and GUs are obtained based on the simulation results of a popular aerospace simulation software Systems Tool Kit (STK) \cite{STK}. 
And the MATLAB 2021b software is used to simulate and assess the performance of the proposed algorithms and schemes for the downlink transmission.
Here, the simulation parameters and constellation configurations are specified at first, and then we present the simulation results and analysis.

\begin{table}[tb]
    \arrayrulecolor{black}
    \centering
    \caption{Simulation Parameters}
    \label{parameter}
    \begin{threeparttable}
    \begin{tabular}{c|c}
        \hline
        Parameter &  Value\\
        \hline
        Orbital height & $h = 1200$ km (LEO) \cite{3GPP-38821}\\
        Minimum elevation angle for GU\tnote{1} & $\theta_{\text{min}} = 10^\circ$\\
        Number of sub-arrays & $N_{x}^\text{sub}$= 8,$N_{y}^\text{sub}$= 4\\
        Number of antenna elements per sub-array & $N_x$ = $N_y$ = 8\\
        Number of GUs & $N_u = 80$\\
        \hline
        Downlink carrier frequency & $f_c=20$ GHz \cite{3GPP-38821}\\
        Bandwidth & 400 MHz \cite{3GPP-38821}\\
        Receiver noise temperature & 24 dBK \cite{hybrid}\\
        Satellite antenna gain\tnote{2} & $G_\text{S} = 21.5$ dBi\\
        GU maximum antenna gain & $G_\text{max} = 40$ dBi \cite{3GPP-38821}\\
        Transmission power per satellite & $P_\text{T}=80$ W\\
        \hline
    \end{tabular}
    \begin{tablenotes}
        \item[1] The minimum elevation angle for a GU is set to guarantee a high LoS probability of the satellite-GU link.
        \item[2] The satellite antenna is a kind of phased array antenna, whose gain can be calculated according to the number and size of antenna elements and the carrier wavelength.
    \end{tablenotes}
    \end{threeparttable}
    % \vspace{-0.5cm}
\end{table}

\subsection{Simulation Parameters}
As described in Section \ref{sec-2}, the multi-satellite cooperative network is operating in the Ka-band with FFR adopted and the satellite antenna array adopts a hybrid architecture where antenna elements are separated into sub-arrays with one RF chain for each sub-array. The specific simulation parameters are given in Table \ref{parameter}.
Throughout the simulation, we use STK to simulate the movement of the LEO satellite constellation and we sample every hour within September 1, 2022 in Beijing time, resulting in 24 experiments to test the performances of the proposed algorithms and schemes. Besides, all GUs are located in suburban areas of 80 representative places in China, which lie between the equator and $54^\circ$ north latitude.

\subsection{Constellation Configuration}
In our study, we analyze the performance of the multi-satellite cooperative network in LEO Walker delta constellations \cite{walker}. 
A Walker delta constellation has all satellites in circular orbits at the same altitude and the same inclination. 
All orbital planes are spaced evenly around the equator and each plane has the same number of satellites which are also spaced evenly in the plane. 
The relative phasing between satellites in adjacent planes is the same for all planes \cite{walker-intro}. 
Herein, we use $T/P/F$ to denote a Walker delta constellation, where $T$, $P$, and $F$ indicate the total number of satellites, the number of orbital planes, and the relative phasing between satellites in adjacent planes, respectively. 
The relative phasing between satellites in adjacent planes is equal to $\frac{2\pi F}{T}$ (rad). 

\begin{figure}[t]
    \centering
    \includegraphics[width=0.48\textwidth]{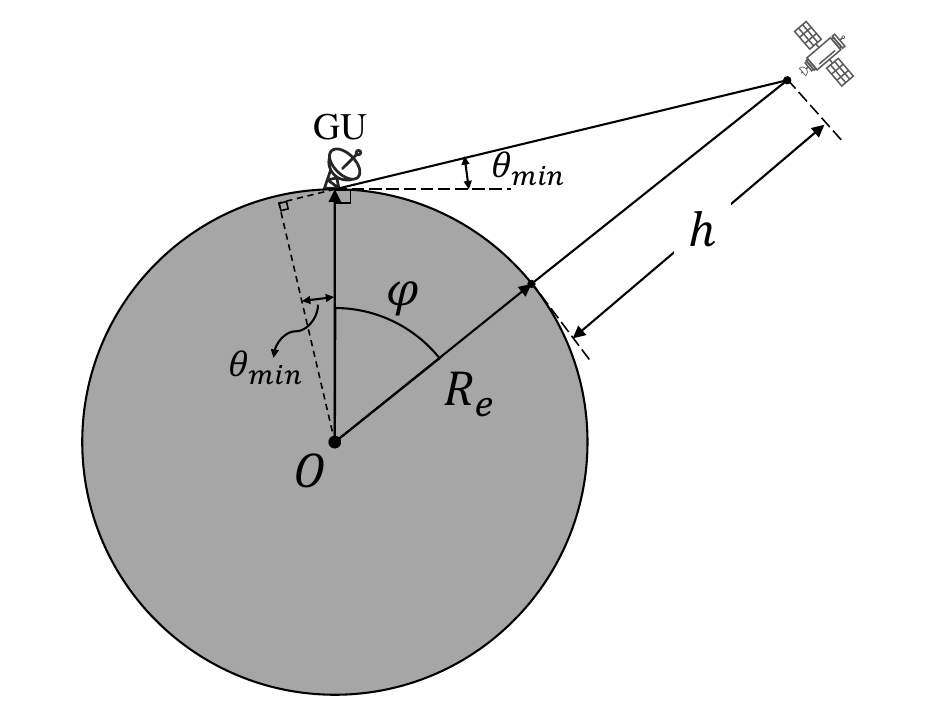}
    \caption{Diagram for calculating the coverage angle according to the orbital height of the satellite and the minimum elevation angle corresponding to the GU.}
    \label{geometry}
    % \vspace{-0.5cm}
\end{figure}

The coverage area of a satellite can be calculated through geometric relations, as illustrated in Fig. \ref{geometry}. 
\begin{figure}[tb]
    \centering
    \includegraphics[width=0.48\textwidth]{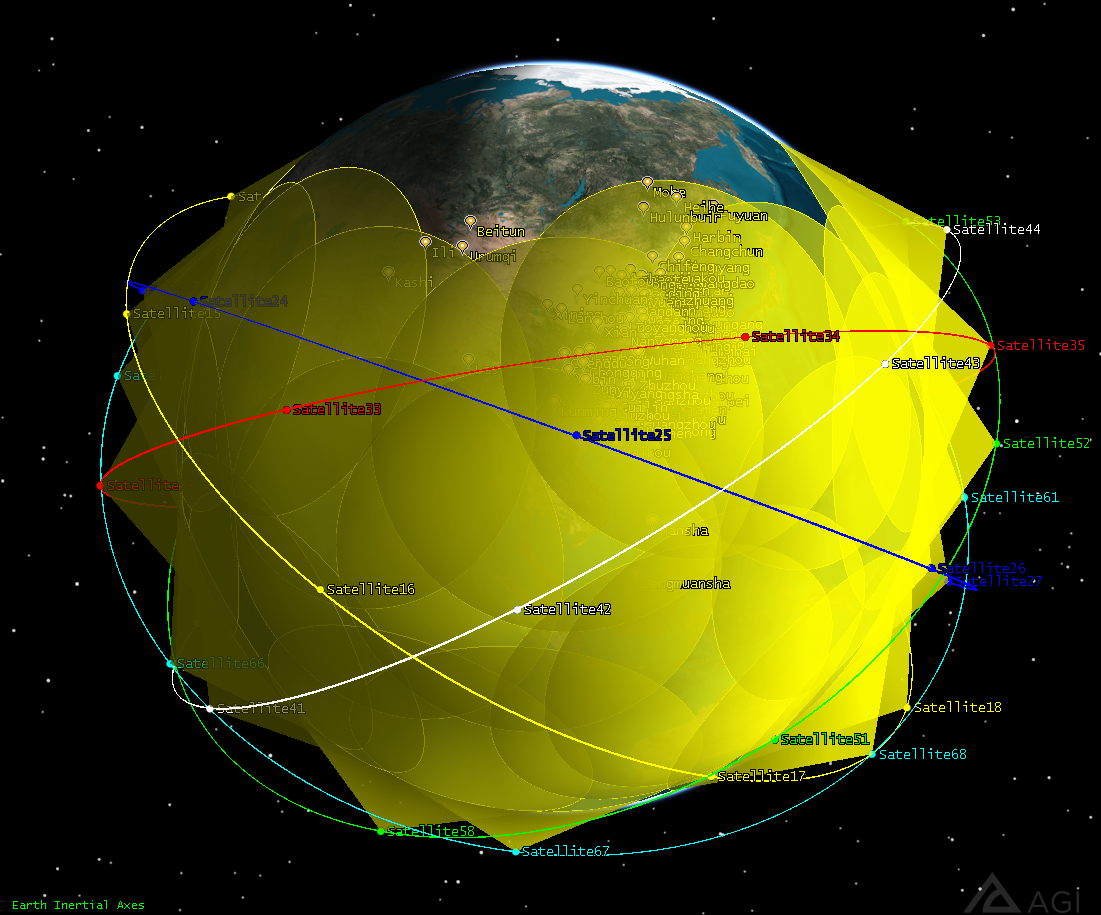}
    \caption{Coverage of a $48/6/1$ constellation with inclination of $30^\circ$.}
    \label{coverage}
    % \vspace{-0.5cm}
\end{figure}
The earth is assumed to be a perfect sphere with center $O$ and radius $R_e$. The satellite is located at orbital height $h$ and $\theta_{\text{min}}$ indicates the minimum elevation angle with respect to GU. 
The relation between the coverage angle $\varphi$, the orbital height $h$, and the minimum elevation angle $\theta_{\text{min}}$ is given by \cite{coverage-angle}
\begin{equation}
    \varphi = \arccos \left( \frac{R_e}{R_e + h} \cos{\theta_\text{min}} \right) -  \theta_\text{min}.
\end{equation}
When $h = 1200$ km and $\theta_\text{min} = 10^{\circ}$, the coverage angle $\varphi$ is approximately equal to $24^{\circ}$. Considering this, the constellation configuration in our study satisfies $P \geq 6$ and $\frac{T}{P} \geq 8$ to realize continuous coverage as much as possible. As mentioned above, all 80 GUs are located between the equator and $54^\circ$ north latitude, which means that the critical inclination of the constellation in our work is $30^\circ$ when $\varphi = 24^\circ$. The coverage of the constellation with critical configuration, i.e., a $48/6/1$ constellation with inclination of $30^\circ$, is illustrated in Fig \ref{coverage}.
It can be seen that most of the GUs are under the coverage of this constellation while several GUs located in high-latitude regions may not be covered in some cases, such as the Beitun city in Fig \ref{coverage}. As such, the inclination of the constellation should be chosen carefully. In our work, we set the inclination in the range of $30^\circ$ to $60^\circ$ and analyze their performance.

\begin{table}[tb]
    \centering
    \caption{Coverage Ratio \& Service Ratio}
    \label{table:ratio}
    \begin{tabular}{c|c|c}
        \hline
        Inclination & Coverage Ratio & Service Ratio\\
        \hline
        $30^\circ$ & $97.14\%$ & $95.99\%$\\
        $35^\circ$ & $99.11\%$ & $98.78\%$\\
        $40^\circ$ & $99.90\%$ & $99.84\%$\\
        $45^\circ$ & $100\%$ & $99.82\%$\\
        $50^\circ$ & $100\%$ & $99.95\%$\\
        $55^\circ$ & $100\%$ & $99.56\%$\\
        $60^\circ$ & $100\%$ & $99.45\%$\\
        \hline
    \end{tabular}
    % \vspace{-0.5cm}
\end{table}
\begin{figure}[tb]
    \centering
    \includegraphics[width=0.4\textwidth]{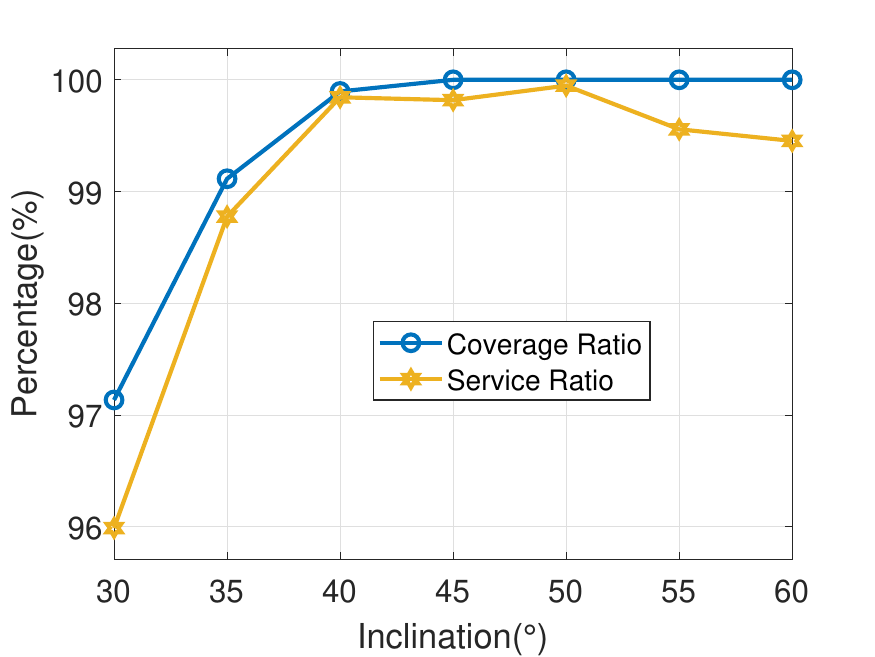}
    \caption{Coverage ratios and service ratios for different inclinations.}
    \label{fig:ratio}
    % \vspace{-0.5cm}
\end{figure}

\subsection{Performance Analysis}
Appropriate representative schemes are not found in the existing literature, so we set the baseline scheme as analog beamforming and user scheduling scheme, denoted as AU, which lacks the procedure of digital beamforming.
The specific steps of AU are: 
(i) analog beamforming as described in Section \ref{sec-3A}, (ii) user scheduling using Algorithm \ref{PUS}, and (iii) power scaling.
In AU and SHU, the whole process of user scheduling is performed based on a fixed beamforming matrix, and the total SE almost reaches its maximum through single-connection procedure. The improvement of multi-connection scheme compared with single-connection scheme is not obvious in AU and SHU.
Therefore, in the sequel, we will compare the performances of the following four schemes: single-connection AU (AU), single-connection SHU (SHU), single-connection JHU (S-JHU), and multi-connection JHU (M-JHU).

\subsubsection{Inclination Selection}

First of all, the choice of constellation inclination is influential to the performances of the multi-satellite cooperative network, and we utilize three indicators to evaluate them: coverage ratio, service ratio, and the mean total SE for a given number of experiments (24 in our simulation). The coverage ratio refers to the ratio of the number of covered GUs to the total number of GUs ($24 \times 80$) in 24 experiments.
Similarly, the service ratio refers to the ratio of the number of served GUs to the total number of GUs ($24 \times 80$) in 24 experiments. The coverage ratios and service ratios of a $48/6/1$ constellation with different inclinations are given in Table \ref{table:ratio} and Fig \ref{fig:ratio}. % Table \ref{tab:ratio} and Fig \ref{ratio}.
We can observe that, without exception, the service ratio is lower than the coverage ratio for each inclination. It is an intuitive result because the satellite can only serve the GUs under its coverage. When the inclination is greater than or equal to $45^\circ$, the coverage ratio can reach 100\%. The service ratio can reach a level very close to 100\% when the inclination is greater than or equal to $40^\circ$ and it shows a tendency of degradation when the inclination increases.

\begin{figure*}[tb]
    \centering
    \includegraphics[width=0.9\textwidth]{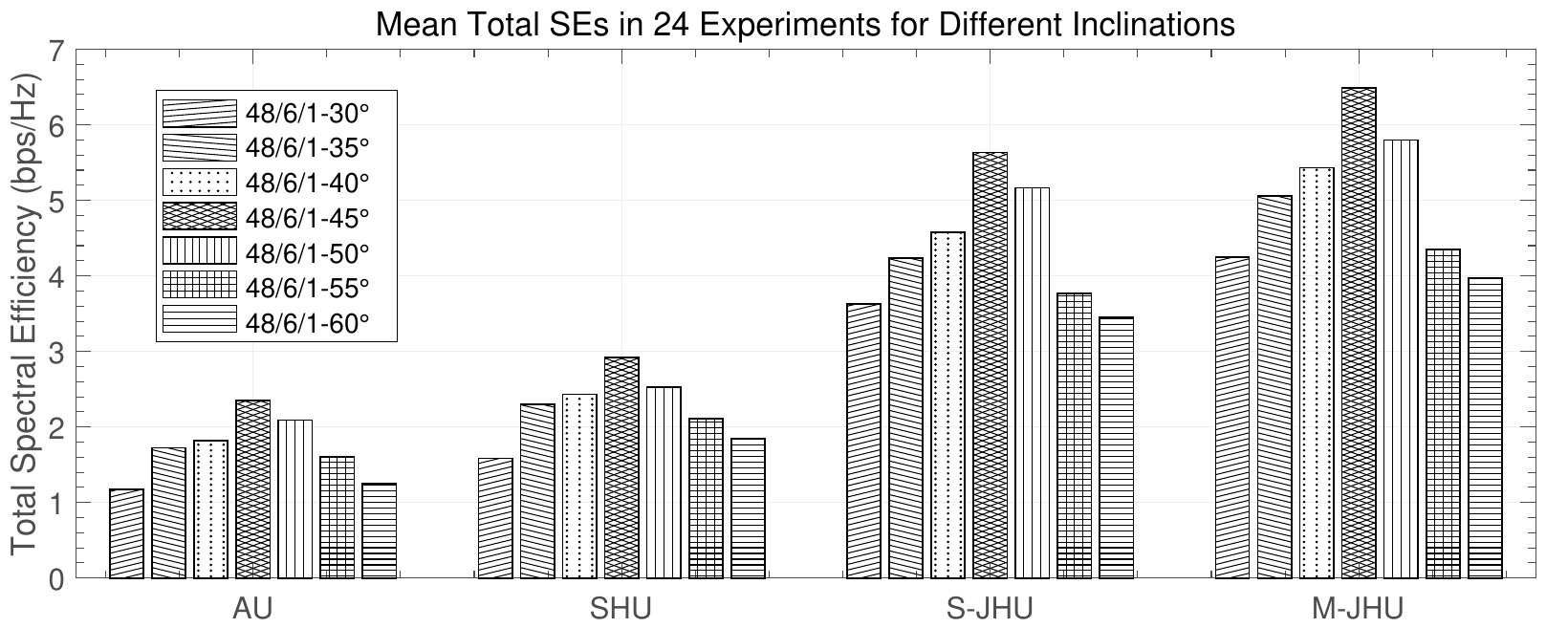}
    \caption{Mean total SEs in 24 experiments for different inclinations of four schemes.}
    \label{Inclination}
    % \vspace{-0.5cm}
\end{figure*}

We also simulate the mean total SEs in 24 experiments for these four schemes of different inclinations, as shown in Fig \ref{Inclination}. The SE performance and the inclination selection appear to be directly related. With the augmentation of the inclination, the mean total SE shows a trend of increasing first and then decreasing, approaching the peak near $45^\circ$ inclination, and the performance seems relatively poor when the inclination is too small or too large. This phenomenon is consistent regardless of the beamforming and user scheduling algorithms. When the inclination is too small, full coverage for GUs may not be possible. GUs located in high-latitude regions are always in the circumstances of low elevation angle, resulting in larger path loss and worse QoS. When the inclination is too large, the number of visible satellites will decline for the low-latitude areas with intensive GUs and some of the visible satellites will cover a large area of high-latitude regions without GUs. Hence, on-board resources are wasted and the system performance degrades.

\begin{figure*}[tb]
    \centering
    \includegraphics[width=0.9\textwidth]{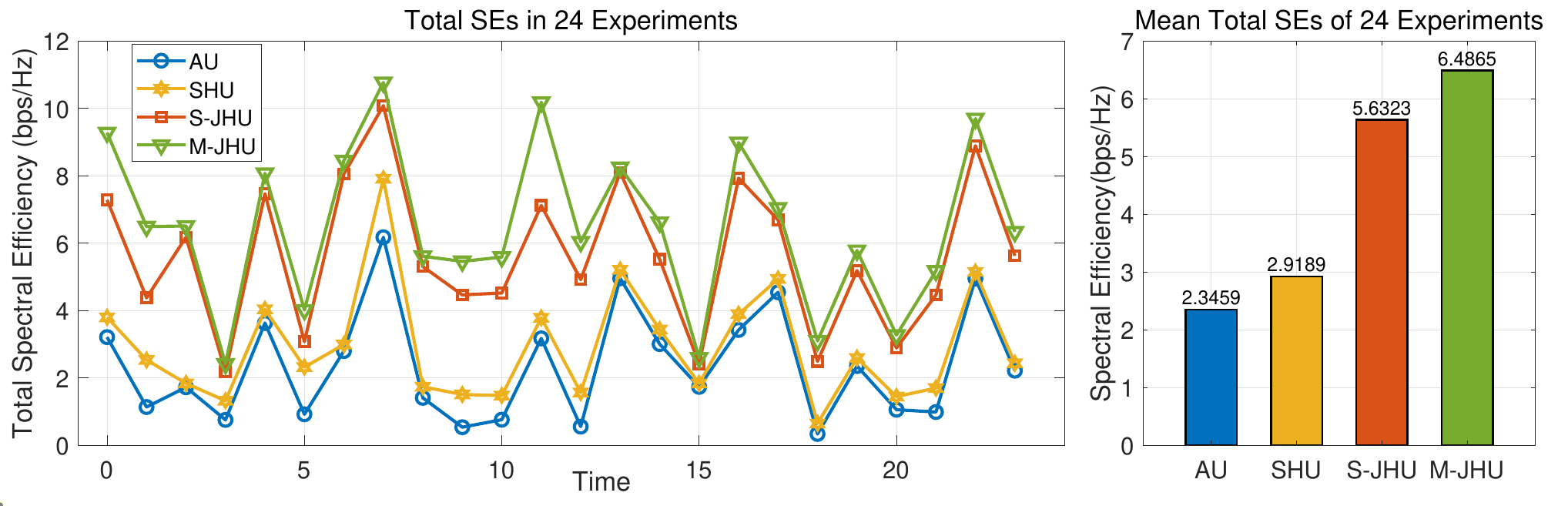}
    \caption{Total SEs and mean total SEs of a $48/6/1$ constellation with inclination of $45^\circ$.}
    \label{sum-mean-48-45}
    % \vspace{-0.5cm}
\end{figure*}
\begin{figure*}[tb]
    \centering
    \includegraphics[width=0.9\textwidth]{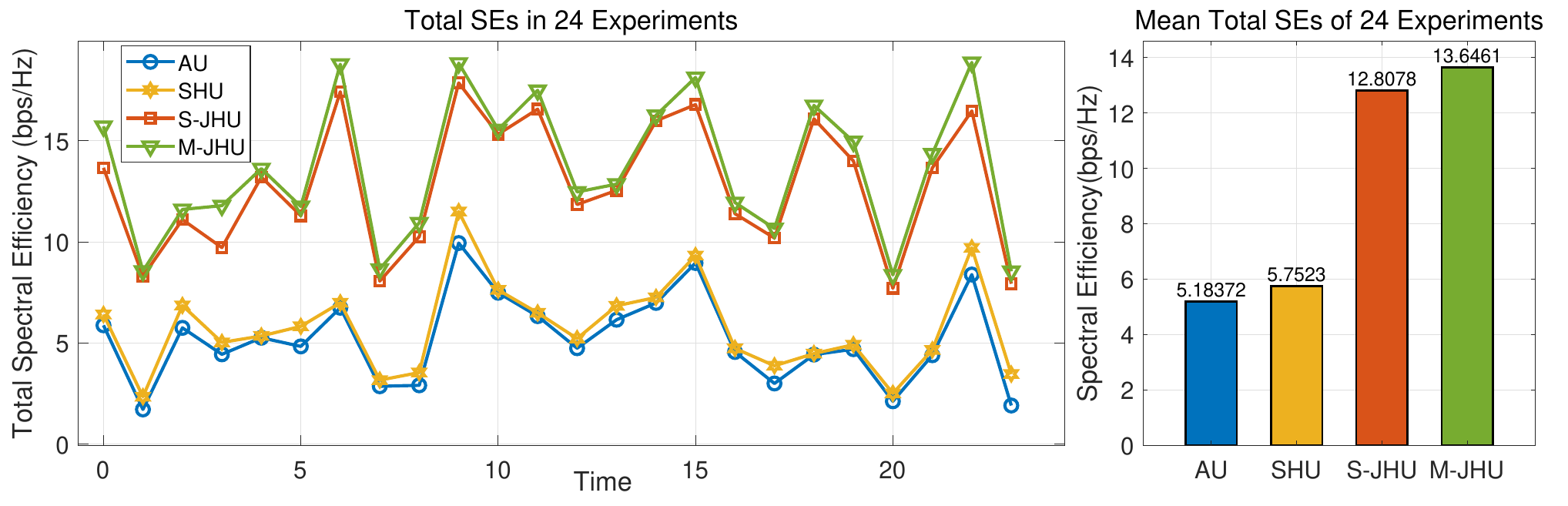}
    \caption{Total SEs and mean total SEs of a $96/6/1$ constellation with inclination of $45^\circ$.}
    \label{sum-mean-96-45}
    % \vspace{-0.5cm}
\end{figure*}
\begin{figure*}[tb]
    \centering
    \includegraphics[width=0.9\textwidth]{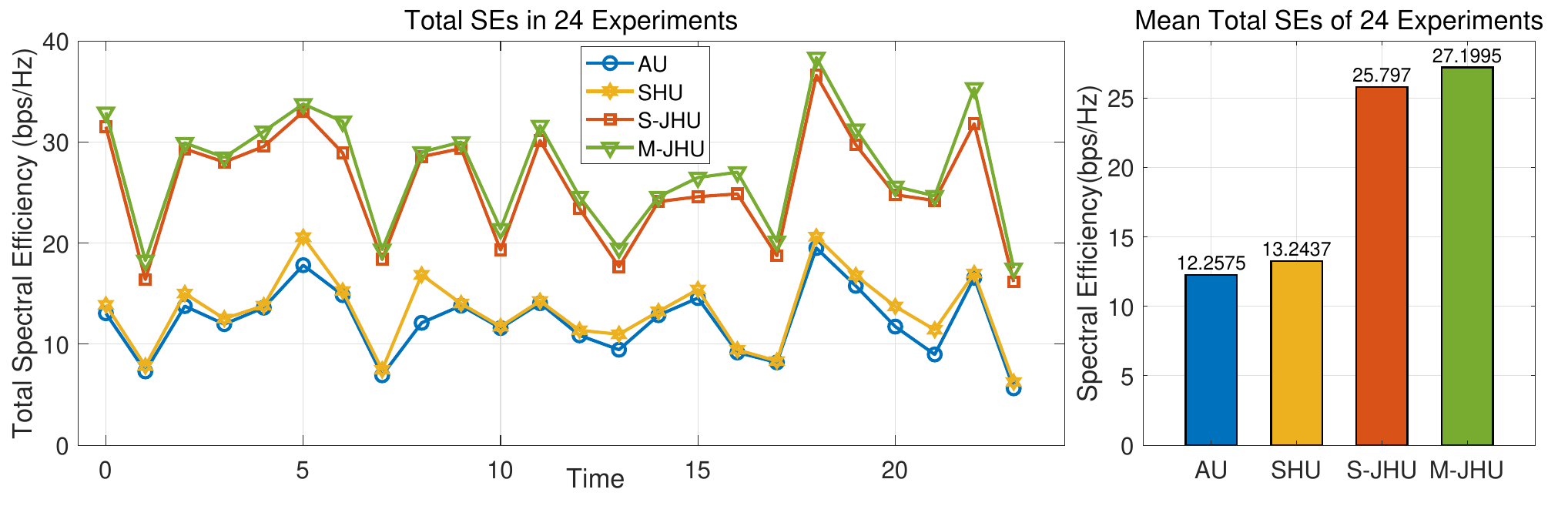}
    \caption{Total SEs and mean total SEs of a $192/12/1$ constellation with inclination of $45^\circ$.}
    \label{sum-mean-192-45}
    % \vspace{-0.5cm}
\end{figure*}

\subsubsection{Algorithm Performance Analysis}
According to the simulation results in the previous part, inclination of $45^\circ$ is adopted in this part. We provide simulation results of three different constellations to illustrate the algorithm performance, as shown in Figs. \ref{sum-mean-48-45}--\ref{sum-mean-192-45}.
For each figure, the left picture illustrates the total SEs of AU, SHU, S-JHU and M-JHU in 24 experiments, and the right picture shows the corresponding mean total SEs averaged over the 24 experiments. 
It can be seen that AU performs the worst in all three constellations because of the lack of interference mitigation. By performing digital beamforming once after user scheduling to mitigate interference, the performance of SHU increases slightly compared with AU on average. However, SHU can not make full use of link information since digital beamforming is implemented independently of user scheduling. In S-JHU and M-JHU, the digital beamforming matrix is updated in real time when calculating the total SE increment and establishing links, leading to a significant increase of SE compared with SHU and AU. In contrast to S-JHU, M-JHU scheme allows one GU to connect with multiple satellites, which can further improve the network SE performance. 
\begin{figure*}[tb]
    \centering
    \begin{minipage}[t]{0.48\textwidth}
        \centering
        \includegraphics[width=\textwidth]{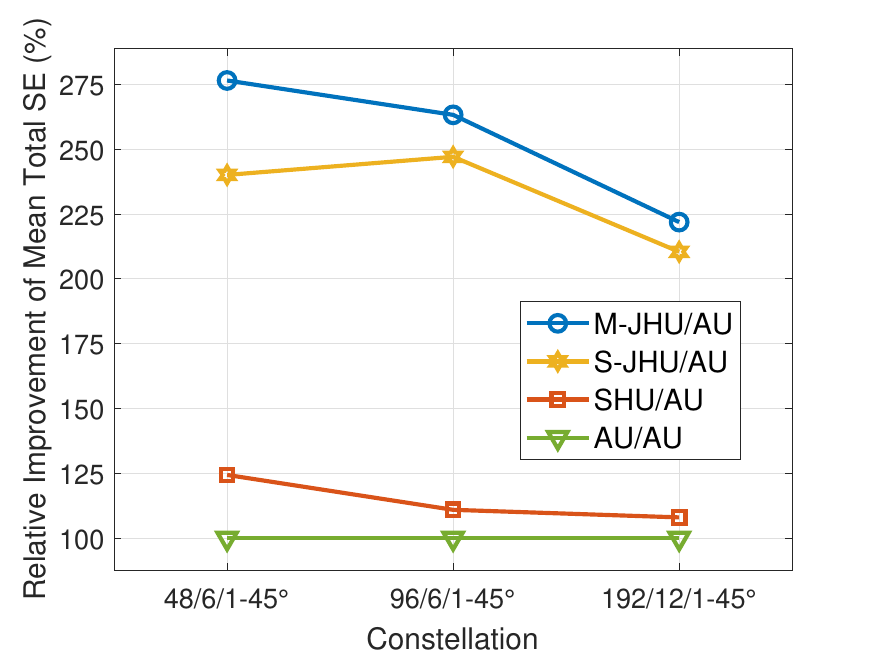}
        \caption{Relative improvement of mean total SE of SHU, S-JHU, and M-JHU over AU in different constellations.}
        \label{gap}
    \end{minipage}
    \begin{minipage}[t]{0.48\textwidth}
        \centering
        \includegraphics[width=\textwidth]{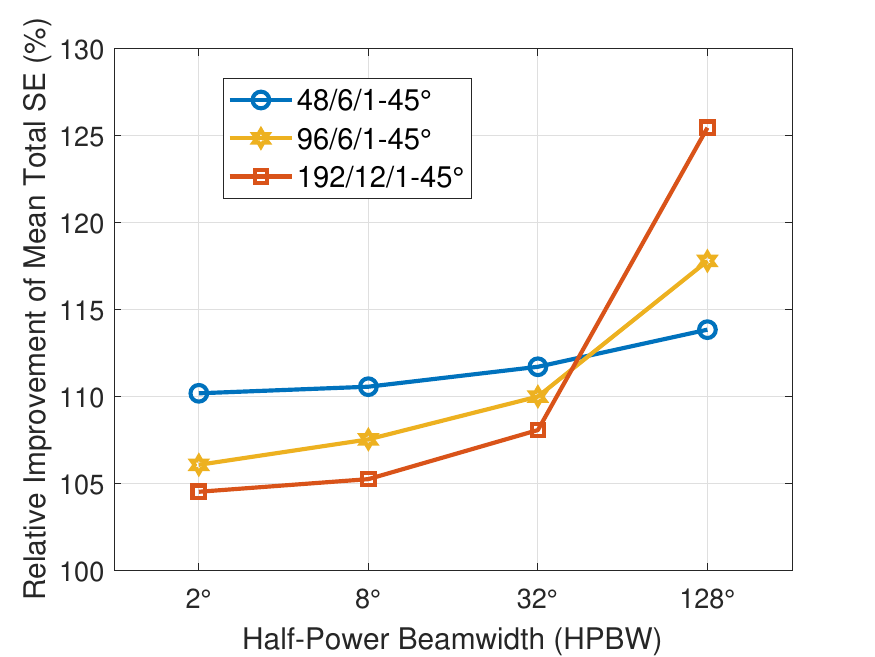}
        \caption{Relative improvement of mean total SE of M-JHU over S-JHU for different GU antenna HPBWs.}
        \label{HPBW}
    \end{minipage}
    % \vspace{-0.5cm}
\end{figure*}

\begin{figure*}[tb]
    \centering
    \subfigure[Beijing]{
        \centering
        \includegraphics[width=0.26\textwidth]{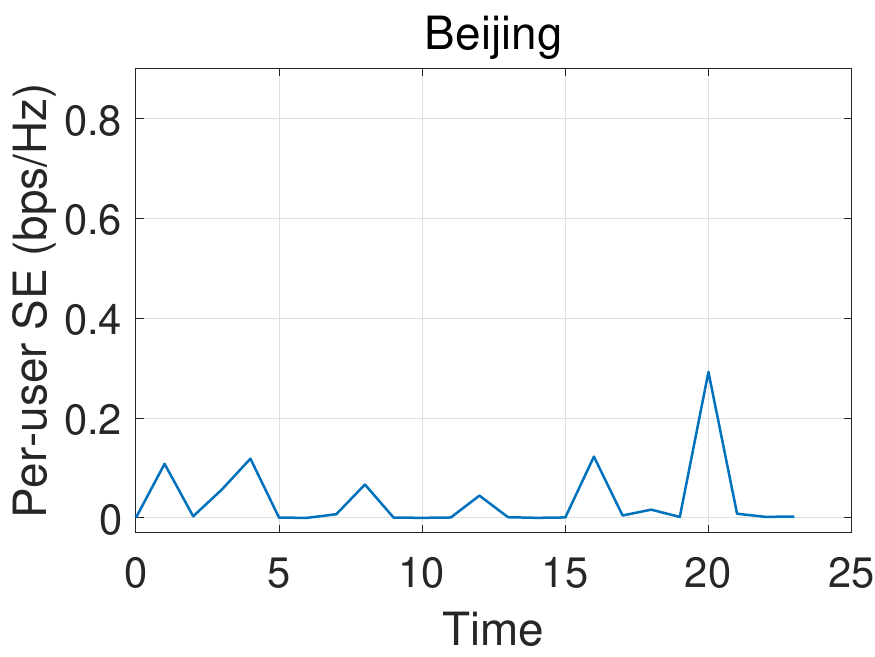}
    }
    \subfigure[Shanghai]{
        \centering
        \includegraphics[width=0.26\textwidth]{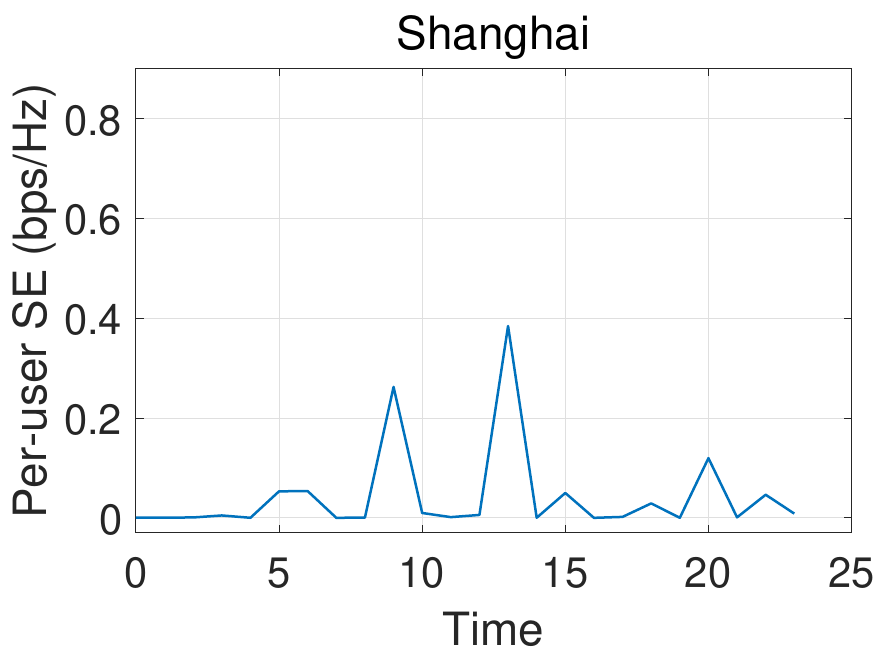}
    }
    \subfigure[Wuhan]{
        \centering
        \includegraphics[width=0.26\textwidth]{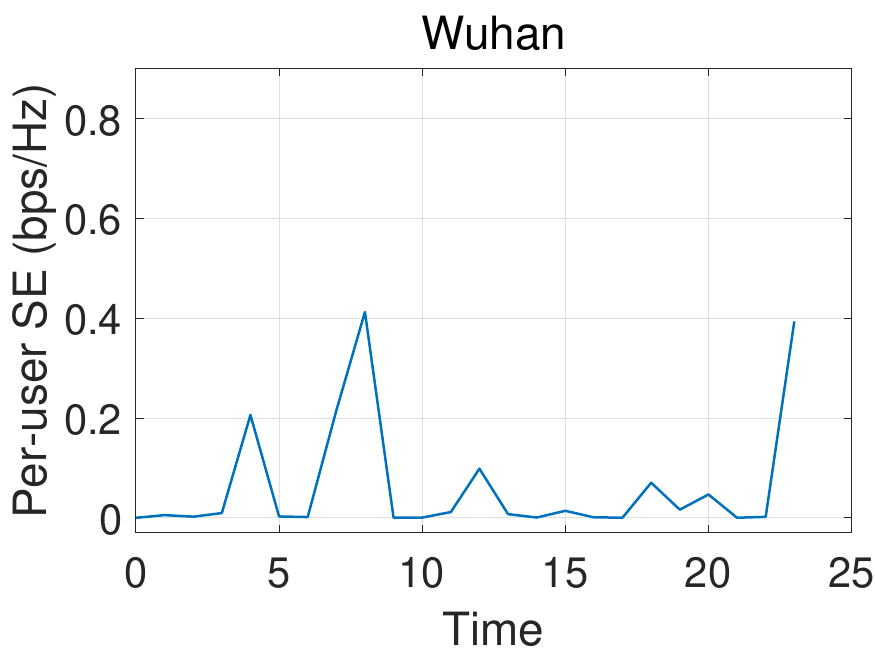}
    }
    \subfigure[Kashi]{
        \centering
        \includegraphics[width=0.26\textwidth]{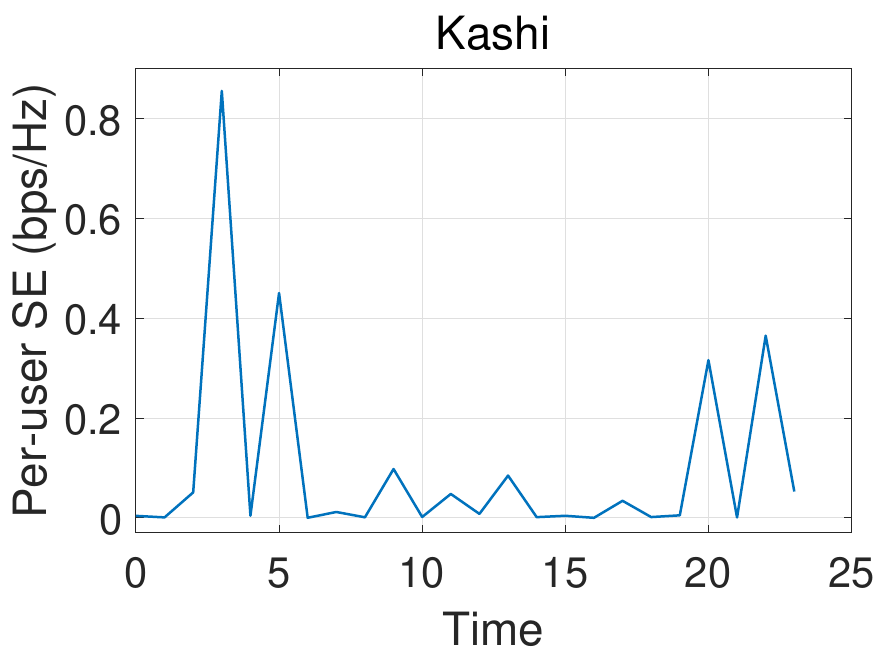}
    }
    \subfigure[Nansha]{
        \centering
        \includegraphics[width=0.26\textwidth]{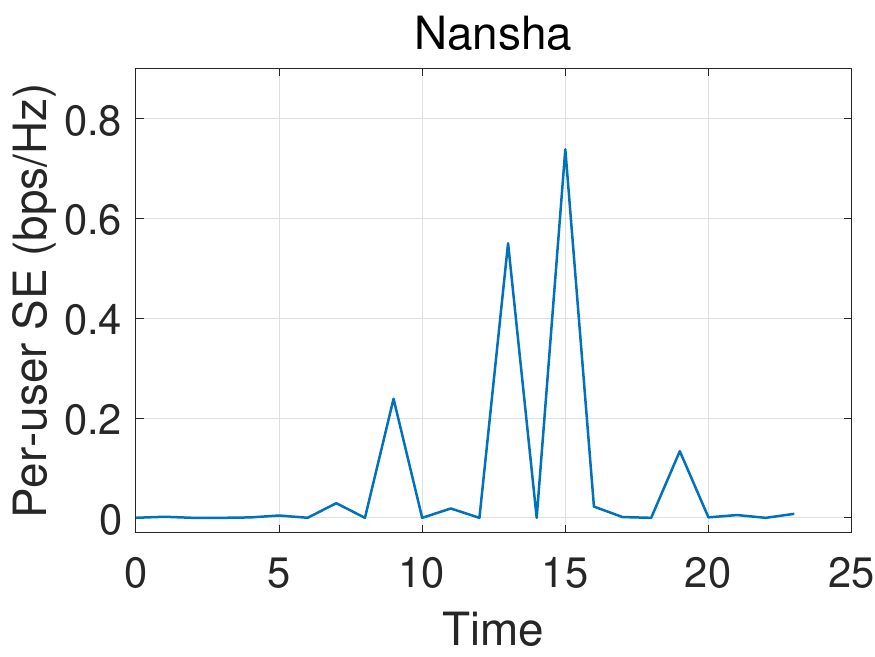}
    }
    \caption{Per-user SEs of GUs in five typical places.}
    \label{5cities}
    % \vspace{-0.5cm}
\end{figure*}

We compare the performance gap of different algorithms in various constellations in Fig. \ref{gap}. It can be seen that the mean total SE improvement of SHU over AU decreases from 24.4\% in the $48/6/1$ constellation to 8\% in the $192/12/1$ constellation. 
The digital beamforming of SHU superior to AU aims at mitigating the inter-beam interference within one satellite. 
When there are more visible satellites, the number of GUs connected to a given satellite decreases hence leading to less inter-beam interference within one satellite which is largely due to the narrow half-power beamwidth (HPBW) of the VSAT at each GU. 
Thus, the effect of digital beamforming is reduced when there are more satellites. 
For similar reasons, compared with AU, the mean total SE improvements of S-JHU and M-JHU are basically reduced as the number of satellites grows.
From Fig. \ref{gap}, we can also see that S-JHU and M-JHU can achieve more than twice mean total SE against AU in different constellations, benefiting from the joint design of user scheduling and beamforming. 
The performance of M-JHU is improved compared with S-JHU in all these three constellations. However, the gap between them becomes smaller and the improvement of M-JHU over S-JHU degrades as the constellation scale increases. 
It means that GUs are more likely to achieve their best SEs through single-connection scheme when the number of visible satellites increases, but multi-connection scheme still has its advantages in small-scale constellations.

The improvement of multi-connection scheme compared with single-connection scheme is also affected by the HPBW of the GU antenna, where the HPBW is twice  $\gamma_\text{3dB}$. 
As the GU antenna HPBW becomes wider, the maximum antenna gain decreases correspondingly. The aforementioned simulations are all executed when the HPBW of the GU antenna is fixed at $1.7^\circ$ and thus the maximum antenna gain is fixed as 40dBi.
If the antenna HPBW changes, the performance will also change. We make a comparison between M-JHU and S-JHU when the HPBW is $2^\circ$, $8^\circ$, $32^\circ$, and $128^\circ$, respectively, as shown in Fig. \ref{HPBW}. 
For the $48/6/1$ constellation, the improvement of M-JHU compared with S-JHU does not change much because the constellation is sparse and there are not many satellites to choose from. While for the $96/6/1$ constellation and $192/6/1$ constellation, there are more satellites to select from and the improvement of M-JHU increases significantly with the augmentation of the HPBW. 
When the HPBW is large enough, for example $128^\circ$ in Fig. \ref{HPBW}, the $192/6/1$ constellation achieves the highest improvement of M-JHU over S-JHU among the three constellations due to the fact that it possesses many more visible satellites for GUs than the other two constellations and the inter-satellite distance is shorter which brings more choices for GUs to establish multiple connections.
Although the multi-connection scheme can achieve more significant improvement compared with the single-connection counterpart at a wider HPBW of the GU antenna, the low received signal power and low SINR brought by the limited antenna gain is still a problem and need to be compensated. In the existing satellite communication systems, high-gain antennas which have narrow HPBW are still the mainstream of non-handheld terminals.

\subsubsection{GU Performance Analysis}

\begin{table}[tb]
    \centering
    \caption{SE statistics of dense and sparse GUs}
    \begin{tabular}{c|c|c}
        \hline
          & Mean Value of SEs (bps/Hz) & Variance of SEs (bps/Hz)$^2$\\
        \hline
        Dense GUs & 0.0732 & 0.1524\\
        Sparse GUs & 0.1048 & 0.3551\\
        \hline
    \end{tabular}
    \label{mean-var}
    % \vspace{-0.5cm}
\end{table}

From Figs. \ref{sum-mean-48-45}--\ref{sum-mean-192-45}, we can observe that the network total SE fluctuates significantly with time. 
The reason is that the topological relationships between satellites and GUs change rapidly with time. Taking GUs in five typical places as examples, we calculate their SEs with M-JHU in a $48/6/1-45^\circ$ constellation and show the results in Fig. \ref{5cities}. 
These five places lie in the northern, eastern, middle, western, and southern part of China, respectively. The SEs of GUs in these five places fluctuate differently due to their distinct geographical positions and topological relationships. 
For the GU in each city, the SE varies with time because of the change of topological relationship, which is caused by satellites' movement. 
These five GUs have very different SEs at the same time because of different sets of visible satellites for these GUs and different elevation angles and distances between satellites and GUs, leading to different path losses. 
The different sets of visible satellites and path losses are both caused by discrepant geographical positions of these GUs. 

In Fig. \ref{5cities}, we can also find that the maximum SEs of GUs in Kashi and Nansha are higher than those in the other three places. Part of the reason lies in the density of GUs. 
A threshold of the inter-GU distance $D$ can divide all GUs into two parts: dense GUs and sparse GUs. Sparse GUs refer to the users in an area where inter-GU distances are all greater than $D$. The opposite holds for dense GUs. For example, we set $D$ as 400 km according to the geographical positions of these 80 GUs in our study. 
With this threshold, the sparse GUs are basically distributed in border areas and their sparsity is very remarkable compared with dense GUs.
Among these 80 GUs, there are 12 sparse GUs and 68 dense GUs. Correspondingly, the GUs in Kashi and Nansha are sparse GUs and the other three GUs are dense GUs. In order to obtain more comprehensive observations, we calculate the SEs of all dense and sparse GUs in 24 experiments and obtain their respective mean value and variance in Table \ref{mean-var}. 
The mean SE of sparse GUs is 43.2\% greater than that of dense GUs. The potential reasons are: The sparse GUs are surrounded by fewer GUs and suffer less interference from others hence having higher SINR; Most of the sparse GUs are located in the border areas of China, thus they have greater probability of monopolizing one satellite and getting larger transmission power. Additionally, the variance of SE of sparse GUs is obviously larger than that of dense GUs, which indicates that the SEs of sparse GUs fluctuate more substantially than dense GUs due to their geographical positions.
If the threshold varies, the difference between dense GUs and sparse GUs will be smaller, but the similar conclusion still holds. For example, if a smaller threshold is chosen, there will be fewer dense GUs and more sparse GUs, and the mean SE of sparse GUs will decrease while the mean SE of dense GUs will remain approximately unchanged. Hence the statistic SE difference between dense GUs and sparse GUs will become smaller, but the mean SE of sparse GUs is still higher than that of dense GUs.

\section{Conclusion}
In this paper, we investigate the SE performance of cooperative multi-LEO-satellite networks. 
First, a hybrid beamforming architecture of UPAs is provided which is suitable for satellites because of the limitation of on-board RF chains. 
According to the hybrid architecture, we introduce an analog beamforming method based on the 2D DFT codebook which generates a desired codeword by linearly combining four selected codewords. 
Then digital beamforming in accordance with the regularized ZF is employed to mitigate the inter-beam interference within one satellite and scale the transmission power. 
Moreover, we propose single-connection and multi-connection heuristic user scheduling algorithms to determine the links between satellites and GUs. 
Subsequently, we propound two implementation schemes: a separate scheme and a joint scheme.
Simulation results show that the S-JHU scheme outperforms SHU in different constellations because of the joint design of beamforming and user scheduling. 
The M-JHU scheme can further improve the SE through connecting multiple satellites to one GU and achieve better performance than S-JHU. 
Then we discuss the effect of constellation configurations and antenna patterns on the performance of M-JHU.
Furthermore, based on the simulation results, we analyze the key factors influencing GUs' SEs, including geographical positions, topological relationships, and the density of GUs. 
The value and variation trend of the SE largely depend on GUs' geographical positions and topological relationships with the satellites, and densely-distributed GUs have relatively lower but stabler SEs compared with sparsely-distributed ones.

% \bibliographystyle{IEEEtran}
% \bibliography{ref.bib} 


\end{document}